\documentclass[12pt,twoside]{article}
\usepackage{correl}
\usepackage{epsf,epsfig,feynmp-auto,chapterbib}
\usepackage{amsmath}
\usepackage{amsthm}
\usepackage[tbtags]{mathtools}
\usepackage{amsbsy} 
\usepackage{cancel}
\usepackage{enumerate}
\usepackage{txfonts}
\usepackage[usenames]{color}
\usepackage{framed}
\usepackage{mathdots}
\usepackage{url}
\usepackage{ulem}
\usepackage{simplewick}
\usepackage{listings,xcolor}
\newcommand \beq  {\begin{equation}}
\newcommand \eeq  {\end{equation}}
\newcommand \beg {\begin{equation}}
\newcommand{\en}{\end{equation}}
\newcommand{\pmat}[1]{\begin{pmatrix} #1 \end{pmatrix}}

\newcommand{\gmat}[1]{\left[\begin{matrix}  #1 \end{matrix} \right]}
\newcommand \bea {\begin{eqnarray} }
\newcommand \eea {\end{eqnarray}}

\newcommand{\bR}{{\bf R}}

\newcommand{\Nsites}{{\cal N}_{s} }
\newlength{\figwidth}

\newlength{\shift}
\def\gtappr{{{\lower4pt\hbox{$>$} } \atop \widetilde{ \ \ \ }}}
\def\ltappr{{{\lower4pt\hbox{$<$} } \atop \widetilde{ \ \ \ }}}
\newcounter{theex}
\newcounter{theap}

\newlength{\spear}

\newlength{\fight}
\newcommand{\fg}[3]
{\begin{figure}[tbh]
\centering
\includegraphics[width=\fight]{#1}
\vskip 0.1truein
\caption{{#2}}\label{#3}\end{figure}}
\newcommand{\fgh}[3]
{\begin{figure}[h]
\centering
\includegraphics[width=\fight]{#1}
\caption{{#2}}\label{#3}\end{figure}}
\newcommand{\fgb}[3]
{\begin{figure}[bh]
\centering
\includegraphics[width=\fight]{#1}
\caption{{#2}}\label{#3}\end{figure}}
\newlength{\bxwidth}

\newcommand \boxit[1]{\noindent\marginpar{\mbox{}}
\fbox{\rule{0pt}{2 \baselineskip}\parbox{1.01\textwidth}{#1}}}
\newcommand \sboxit[1]{\vskip 0.2truein\noindent\marginpar{\mbox{}}
\fbox{\rule{0pt}{2 \baselineskip}\parbox{0.98\textwidth}{#1}}
\vskip 0.2truein\noindent}
\newenvironment{Quote}%
  {\begin{list}{}{%
	\setlength{\leftmargin}{0.5truein}
       \setlength{\rightmargin}{0.5\leftmargin}}
           \item[]\ignorespaces}
  {\end{list}}
\newlength{\wx}
\newlength{\wy}
\newlength{\wz}
\def\3he{{$^3${\rm He}}}

\def\slD{\raise.15ex\hbox{$/$}\kern-.57em\hbox{$D$}}
\def\dsl{\raise.15ex\hbox{$/$}\kern-.57em\hbox{$\Delta$}}
\def\slp{{\raise.15ex\hbox{$/$}\kern-.57em\hbox{$\partial$}}}
\def\nsl{\raise.15ex\hbox{$/$}\kern-.57em\hbox{$\nabla$}}
\def\sla{\raise.15ex\hbox{$/$}\kern-.57em\hbox{$\rightarrow$}}
\def\slla{\raise.15ex\hbox{$/$}\kern-.57em\hbox{$\lambda$}}
\def\gtwid{\raise.3ex\hbox{$>$\kern-.75em\lower1ex\hbox{$\sim$}}}
\def\ltwid{\raise.3ex\hbox{$<$\kern-.75em\lower1ex\hbox{$\sim$}}}

\def\12{{1\over2}}

\def\part{\partial}

\def\bk{{\bf k}}

\def\bethlogo{\vbox{\bf \line{\hrulefill} 
    \kern-.5\baselineskip 
    \line{\hrulefill\phantom{ ELIZABETH A. MASON }\hrulefill} 
    \kern-.5\baselineskip 
    \line{\hrulefill\hbox{ ELIZABETH A. MASON }\hrulefill} 
    \kern-.5\baselineskip 
    \line{\hrulefill\phantom{ 1411 Chino Street }\hrulefill} 
    \kern-.5\baselineskip 
    \line{\hrulefill\hbox{ 1411 Chino Street }\hrulefill} 
    \kern-.5\baselineskip 
    \line{\hrulefill\phantom{ Santa Barbara, CA 93101 }\hrulefill} 
    \kern-.5\baselineskip 
    \line{\hrulefill\hbox{ Santa Barbara, CA 93101 }\hrulefill}
    \kern-.5\baselineskip 
    \line{\hrulefill\phantom{ (805) 962-2739 }\hrulefill} 
    \kern-.5\baselineskip 
    \line{\hrulefill\hbox{ (805) 962-2739 }\hrulefill}}}
\def\lisalogo{\vbox{\bf \line{\hrulefill} 
    \kern-.5\baselineskip 
    \line{\hrulefill\phantom{ LISA R. GOODFRIEND }\hrulefill} 
    \kern-.5\baselineskip 
    \line{\hrulefill\hbox{ LISA R. GOODFRIEND }\hrulefill} 
    \kern-.5\baselineskip 
    \line{\hrulefill\phantom{ 6646 Pasado }\hrulefill} 
    \kern-.5\baselineskip 
    \line{\hrulefill\hbox{ 6646 Pasado }\hrulefill} 
    \kern-.5\baselineskip 
    \line{\hrulefill\phantom{ Santa Barbara, CA 93108 }\hrulefill} 
    \kern-.5\baselineskip 
    \line{\hrulefill\hbox{ Santa Barbara, CA 93108 }\hrulefill}
    \kern-.5\baselineskip 
    \line{\hrulefill\phantom{ (805) 962-2739 }\hrulefill} 
    \kern-.5\baselineskip 
    \line{\hrulefill\hbox{ (805) 962-2739 }\hrulefill}}}

\def\low#1{\lower.5ex\hbox{${}_#1$}}
\def\ltwid{\raise.3ex\hbox{$<$\kern-.75em\lower1ex\hbox{$\sim$}}}

\def\psl{\raise.15ex\hbox{$/$}\kern-.57em\hbox{$\partial$}}
\def\partt{\raise.15ex\hbox{$\widetilde$}{\kern-.37em\hbox{$\partial$}}}
\def\parts{\raise.15ex\hbox{$/$}{\kern-.6em\hbox{$\partial$}}}
\def\nablas{\raise.15ex\hbox{$/$}{\kern-.6em\hbox{$\nabla$}}}
\def\oprod{\hbox{$\rm O$}{\kern -0.8em\hbox{$\Pi$}}}
\def\partw#1{\raise.15ex\hbox{$/$}{\kern-.6em\hbox{${#1}$}}}

\def\si{{\sigma}}

\def\gtappr{{{\lower4pt\hbox{$>$} } \atop \widetilde{ \ \ \ }}}
\def\ltappr{{{\lower4pt\hbox{$<$} } \atop \widetilde{ \ \ \ }}}

\def\topppageno1{\global\footline={\hfil}\global\headline
={\ifnum\pageno<\firstpageno{\hfil}\else{\hss\twelverm --\ \folio
\ --\hss}\fi}}

\def\toppageno2{\global\footline={\hfil}\global\headline
={\ifnum\pageno<\firstpageno{\hfil}\else{\rightline{\hfill\hfill
\twelverm \ \folio
\ \hss}}\fi}}

\def\ltdash{\raise-1.8pt\hbox{$\scriptscriptstyle |$}}

\def\dg{{^
{\dag}}}

\def\1{{\bf 1}}
\def\2{{\bf 2}}

\def\ell{{\it l } {\rm n}}

\def\si{\sigma}

\def\cx2{\sqrt{c^2_x+c^2_y}}

\def\gkk{\gamma _{\vec k}}
\def\gk2{\gkk ^2}
\def\dw{\downarrow}
\def\up{\uparrow}
\def\gtappr{{{\lower4pt\hbox{$>$} } \atop \widetilde{ \ \ \ }}}
\def\ltappr{{{\lower4pt\hbox{$<$} } \atop \widetilde{ \ \ \ }}}

\def\pbar{{\partial\kern-1.2ex\raise0.25ex\hbox{/}}}

\def\up{\uparrow}
\def\dw{\downarrow}

\def\dg{{^{\dag}}}

\def\1{{\bf 1}}
\def\2{{\bf 2}}

\def\ell{{\it l } {\rm n}}

\def\si{\sigma}

\def\cx2{\sqrt{c^2_x+c^2_y}}

\def\gkk{\gamma _{\vec k}}
\def\gk2{\gkk ^2}
\def\gtappr{{{\lower4pt\hbox{$>$} } \atop \widetilde{ \ \ \ }}}
\def\ltappr{{{\lower4pt\hbox{$<$} } \atop \widetilde{ \ \ \ }}}

\def\thickra{\hbox{\raise0.2pt\hbox{{$\bf >\mkern-13mu>\mkern-13mu>$}}}}
\def\thickrarrow{\hbox{\raise0.28pt\hbox{{$\bf >\mkern-13mu>\mkern-13mu>$}}}}

\newlength{\upit}\upit=0.1truein

\newcommand \bfyn{\begin{fmfgraph*}}
\newcommand \bfynu{\begin{fmfgraph}}
\newcommand \bfynq{\begin{fmfgraph*}}
\newcommand\bfynt{\begin{fmfgraph*}}
\newcommand \efyn{\end{fmfgraph*}}

\newcommand\bprop[1]
{\parbox{15mm}{\bfyn(15,5)
\fmfleft{i}\fmfright{o}\fmf{dbl_wiggly,label=$\noexpand {#1}$}{i,o}\end{fmfgraph*}}}
\newcommand\wprop[3]{
\fmf{boson,label=$\noexpand {#3}$}
{#1,#2}}
\newcommand\fprop[1]
{\parbox{15mm}{\bfyn(15,5)
\fmfleft{i}\fmfright{o}\fmf{dbl_plain_arrow,label=#1}{i,o}\end{fmfgraph*}}}
\newcommand\fpropt[1]
{\parbox{15mm}{\bfyn(15,5)
\fmfpen{thick} \fmfleft{i}\fmfright{o}\fmf{fermion,label=#1,fore=blue}{i,o}\end{fmfgraph*}}}
\newcommand\doubprop[3]{
\fmf{dbl_plain_arrow,label=$\noexpand {#3}$}
{#1,#2}}
\newcommand\lprop[3]{
\fmf{fermion,label=$\noexpand {#3}$}
{#1,#2}}
\newcommand\opold[2]{\bfyn(#1,#2)
\fmfright{i1,i2}\fmfleft{o1}\fmf{fermion}{i1,o1,i2}\fmfdot{o1}
\fmfv{l=$\noexpand {\delta\over\delta  \bar \alpha (1)}$,l.a=0,l.d=0.1w}{i1}
\fmfv{l=$\zeta\noexpand {\delta\ \over \delta  \alpha (1) }$,l.a=0,l.d=0.1w}{i2}
\efyn
}
\newcommand\opa[2]{\bfyn(#1,#2)
\fmfright{i1,i2}\fmfleft{o1}\fmf{phantom,tension=15}{o1,o2}\fmf{fermion}{i1,o2,i2}\fmfdot{o2}
\fmfv{l=$\noexpand {\delta\over\delta  \bar \alpha}$,l.a=0,l.d=0.1w}{i1}
\fmfv{l=$\zeta\noexpand {\delta\ \over \delta  \alpha  }$,l.a=0,l.d=0.1w}{i2}
\efyn
}
\newcommand\opbb[2]{\bfyn(#1,#2)
\fmfleft{i1,i2}\fmfright{o1,o2}\fmf{fermion}{i1,v1,i2}\fmf{fermion}{o1,v2,o2}
\fmf{boson}{v1,v2}
\fmfv{l=$\noexpand {\delta\over\delta \bar\alpha(2)}$,l.a=180,l.d=0.1w}{i1}
\fmfv{l=$\noexpand {\delta\over\delta \alpha(2)}$,l.a=180,l.d=0.1w}{i2}
\fmfv{l=$\noexpand {\delta\over\delta \bar \alpha(1)}$,l.a=0,l.d=0.1w}{o1}
\fmfv{l=$\noexpand {\delta\over\delta \alpha(1)}$,l.a=0,l.d=0.1w}{o2}
\efyn
}
\newcommand\opb[2]{\bfyn(#1,#2)
\fmfleft{i1,i2}\fmfright{o1,o2}\fmf{fermion}{i1,v1,i2}\fmf{fermion}{o1,v2,o2}
\fmf{boson}{v1,v2}
\fmflabel{$\noexpand {\delta\over\delta \bar\alpha(2)}$}{i1}
\fmflabel{$\noexpand {\delta\over\delta \alpha(2)}$}{i2}
\fmflabel{$\noexpand {\delta\over\delta \bar \alpha(1)}$}{o1}
\fmflabel{$\noexpand {\delta\over\delta \alpha(1)}$}{o2}
\efyn
}

\newcommand\blob[2]{
\fmfv{label=$\noexpand #1$,label.dist=-1mm,
decoration.shape=circle,decoration.size=7mm,decoration.filled=0}{#2}
}

\newcommand{\SMB}{SmB$_6$}

\def\TheTitle{Heavy Fermions and the Kondo Lattice: a 21st Century Perspective}
\def\TheHeading{Heavy Fermions and the Kondo Lattice}
\def\TheAuthor{Piers Coleman}
\def\TheInstitute{Center for Materials Theory, Rutgers University}
\def\TheAddress{136 Frelinghuysen Road, Piscataway NJ 08854, USA}
\def\TheChapter{5}                       
\begin{document}
\MakeTitle           
\begin{fmffile}{feyn/zombybk17}

%
\fight=0.5 \textwidth
\vskip 0.2truein



\section{Heavy  Electrons}\label{}

\subsection{Introduction}\label{}
In a world where it is possible to hold a levitated
high temperature superconductor
in the palm of one's hand, it is easy to forget the ongoing importance
of low temperature research. 
Heavy electron materials are a 
class of strongly correlated electron material\index{strongly
correlated materials}
containing localized magnetic moments\index{local moments} which, by entangling with the
surrounding electrons,  profoundly transform the metallic properties.
A heavy fermion metal\index{heavy fermion metal} can develop electron masses 1000 times that of
copper, it can also develop unconventional superconductivity, 
transform into new forms of quantum order, exhibit 
quantum critical and topological behavior. 
Although most of these properties develop well below
the boiling point of nitrogen, the diversity and highly tunable
nature of their ground-states make them 
an invaluable vital work-horse for exploring and researching the
emergent properties of correlated quantum matter. 

This lecture will give an introduction to heavy fermion
materials\index{heavy fermion materials},
trying to emphasize a 21st century perspective.  More extensive
discussion and development of the ideas in these notes can be found
in an earlier review article \cite{onthebrink} and the latter chapters
of my book ``Introduction to Many Body Physics''\cite{intromanybody}.

\fight=0.65\textwidth
\fg{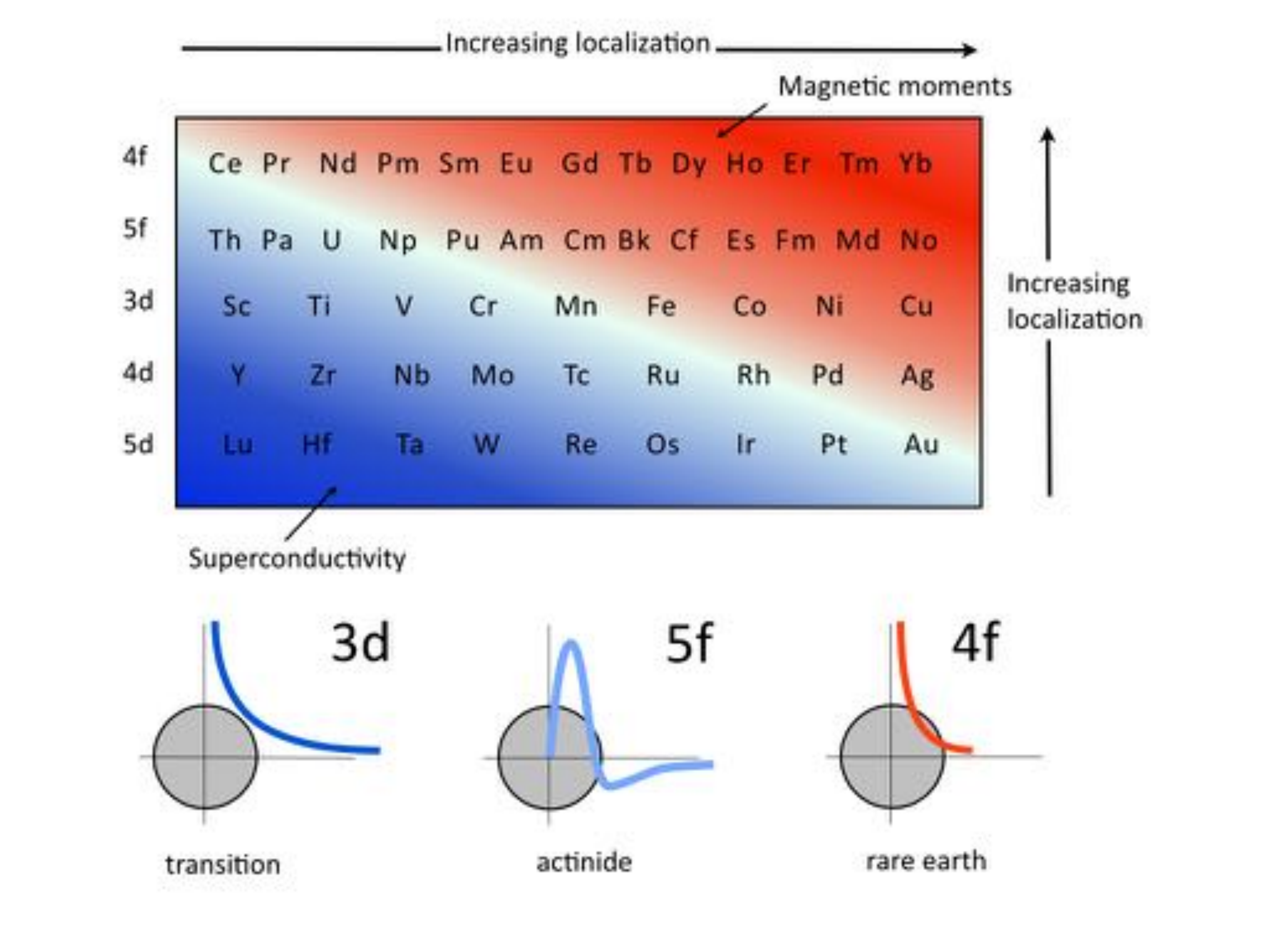}
{The Kmetko-Smith
diagram\cite{kmetkosmith}, showing the broad trends towards increasing
electron localization in the d- and f-electron compounds.  
}{kmetko}

In the periodic table, the most strongly interacting electrons reside 
in orbitals that are well-localized. 
In order of increasing
localization, partially filled orbitals are ordered as follows:
\begin{equation}\label{}
5d < 4d < 3d < 5f < 4f.
\end{equation}
In addition, when moving along a row of the periodic table,
the increasing nuclear charge pulls the orbitals towards the nucleus. 
These trends are summarized in the 
``{Kmetko-Smith} diagram''%
\cite{kmetkosmith} in Fig \ref{kmetko}.
The d-orbital metals at the bottom left of this diagram 
are highly itinerant and 
exhibit conventional superconductivity. 
By contrast, in rare earth and actinide 
metals towards the top right-hand corner, 
the f-shell electrons are localized, forming magnets or antiferromagnets. 
It is the materials  that lie in the 
cross-over between these two regions that are particularly interesting,
for these materials are ``on the brink of magnetism''.
It is in this cross-over region that many 
strongly correlated materials reside:
it is here for instance, that
 we find cerium and uranium, which are key atoms for a wide range of
 4f and 5f heavy electron materials\index{heavy fermion
 materials}\index{heavy electron materials}.

\subsection{Local moments and the Kondo effect}\label{}
Heavy electron materials contain a lattice of localized electrons 
immersed in a sea of mobile conduction electrons.  To understand their
physics, we need to first step back and discuss individual localized
moments\index{local moments}, and the mechanism by which they interact with the surrounding
conduction sea. 

The key feature of a localized moment, is that the Coulomb interaction
has eliminated the high frequency charge fluctuations, leaving behind
a low energy manifold of degenerate spin states. 
In rare earth and actinide ions, the
orbital and spin angular momentum combine into a single entity with 
angular momentum $\vec{j} = \vec{l}+\vec{s}$. 
For example, a $Ce^{3+}$ ion contains a single unpaired 4f-electron in the state
$4f^{1}$, with $l=3$ and $s=1/2$. Spin-orbit coupling gives rise to 
low-lying multiplet with $j = 3-\frac{1}{2}= \frac{5}{2}$, consisting of 
$2j+1=6$ degenerate orbitals $\vert 4f^{1}:J m\rangle $,
($m_{J}\in[-\frac{5}{2},\frac{5}{2}]$) with an associated magnetic
moment $M = 2.64 \mu_{B}$.  In a crystal, the $2j+1$ fold degeneracy
of such a magnetic ion is split, and provided there are an odd number
of electrons in the ion, Kramer's theorem guarantees that the lowest
lying state has at least, a two fold degeneracy. (Fig. \ref{fig2x} a
and b.)

One of the classic signatures of localized
moments,  is a high temperature Curie Weiss susceptibility\index{Curie
Weiss susceptibility}, given by
\begin{equation}\label{curie}
\chi \approx n_{i}  \frac{M^2}{3 (T+\theta )} \qquad \qquad M^2 = g^2\mu _{B}^{2}j(j+1), 
\end{equation}
where, $n_{i}$ is the concentration of magnetic moments while $M$ is the
magnetic moment with 
total angular momentum quantum number $j$ and gyro-magnetic ratio
(``g-factor'') $g$.   $\theta
$ is the ``Curie Weiss'' temperature, a phenomenological 
scale which takes account of interactions between
spins. 

\fight=0.7\textwidth
\fgh{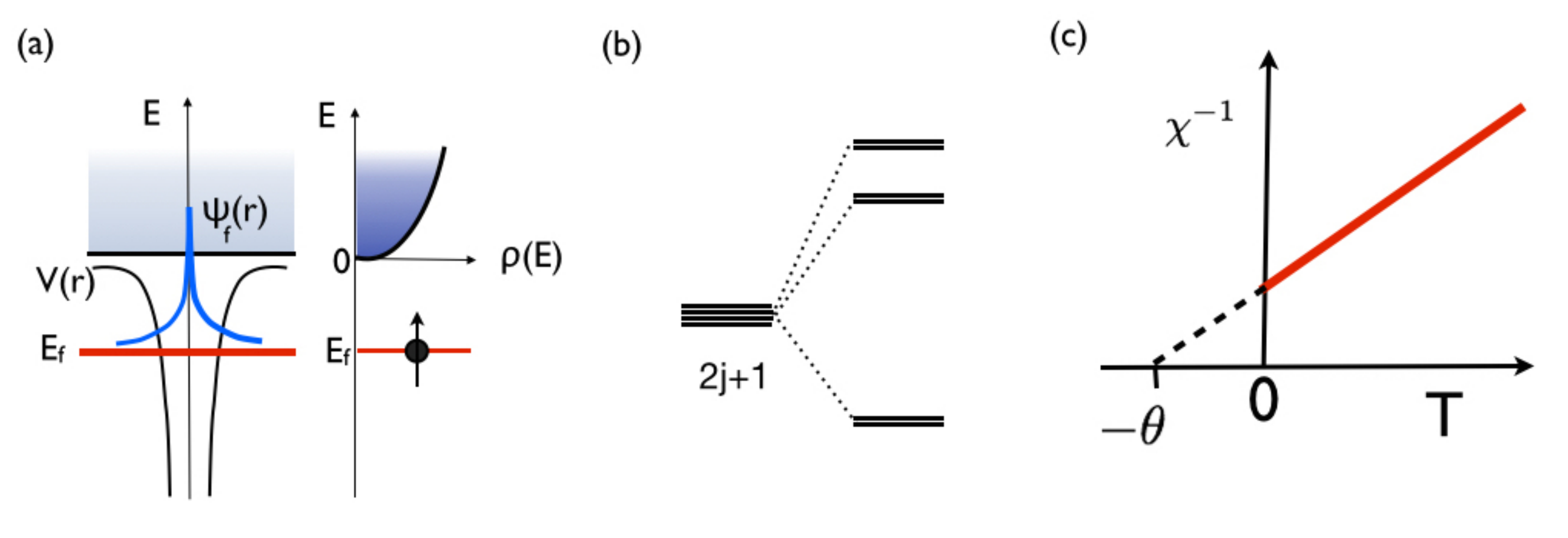}{(a) In isolation, the localized
atomic states of an atom form a stable, sharp excitation lying below the
continuum. (b) In a crystal, the $2j+1$ fold degenerate state splits
into multiplets, typically forming a low lying Kramers doublet.  (c)
The inverse of the Curie-Weiss susceptibility\index{Curie Weiss
susceptibility} of local moments\index{local moments}
$\chi^{-1} $  is a linear function of temperature, intersecting zero
at $T = - \theta $.
}{fig2x} 

The presence of such local moments inside a metal profoundly alters 
its properties. The physics of an isolated magnetic ion is described
by the {\sl Kondo model}\index{Kondo model}
\begin{equation}\label{kondomodel2}
H= \sum_{k\sigma }\epsilon _{k}c\dg _{k\sigma }c_{k\sigma } +\overbrace{J
\psi \dg (0)\vec{\sigma
}\psi (0)\cdot \vec{S}_{f}}^{\Delta H}.
\end{equation}
where $c\dg_{k\sigma }$ creates a conduction electron of energy
$\epsilon_{k}$, momentum $k$  and 
$\psi\dg(0)={\cal N}_{s}^{-1/2}\sum_k c\dg_{k\sigma }$ creates a conduction
at the origin, where ${\cal N}_{s}$ is the number of sites in the lattice. The conduction sea interacts 
with local moment via an antiferromagnetic contact interaction of strength
$J$. 
  The
antiferromagnetic sign ($J>0$) of this interaction is an example of
``super-exchange'', first predicted by Philip
W. Anderson\cite{Anderson:1959bk,anda}, which results
from high energy valence fluctuations. 
Jun Kondo\cite{kondo2} first analyzed the
effect of this scattering, showing that as the temperature is lowered,
the effective strength of the interaction grows logarithmically,  according to 
\begin{equation}\label{}
J \rightarrow J (T)= J + 2 J^{2}\rho \ln \frac{D}{T}
\end{equation}
where $\rho $ is the density of states of the conduction sea (per
spin) and $D$ is the band-width.   The growth of this interaction
enabled Kondo to understand why in many metals at low temperatures,
the resistance starts to rise as the temperature is lowered, giving
rise to  {\sl resistance minimum}. 
\fight=0.95\textwidth
\fg{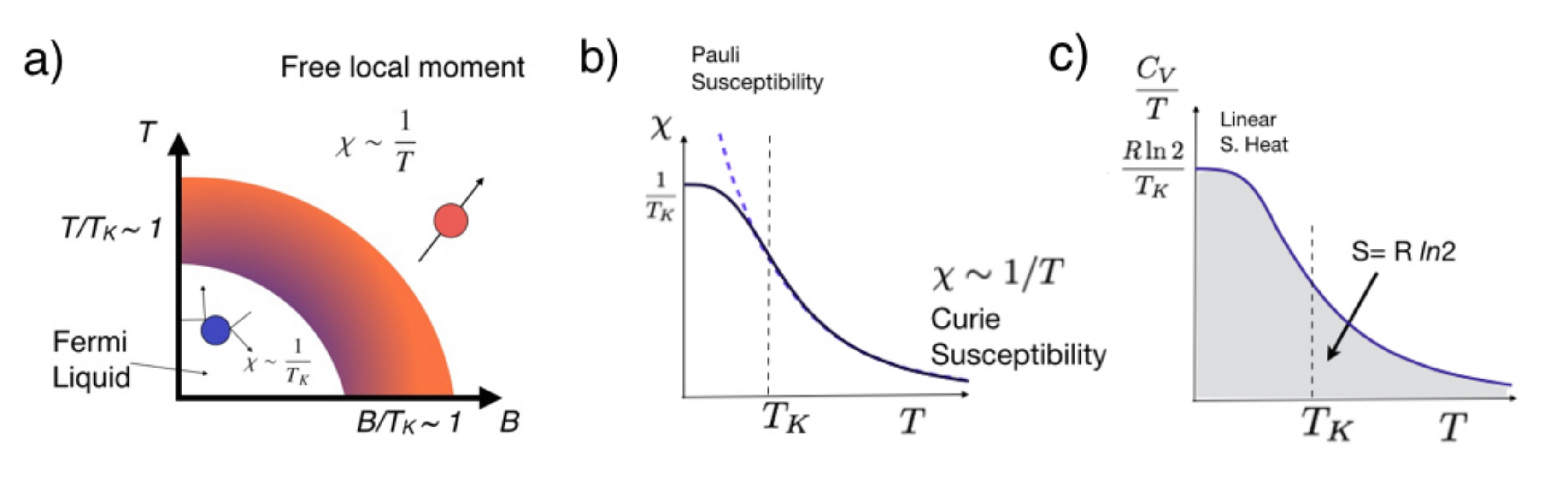}{
(a) Schematic temperature-field phase diagram of the Kondo effect.  At
fields and temperatures large compared with the Kondo temperature
$T_{K}$, the local moment is unscreened with a Curie susceptibility. 
At temperatures and fields
small compared with $T_{K}$, the local moment is screened, forming an
elastic scattering center within a Landau Fermi liquid with a
Pauli susceptibility $\chi \sim \frac{1}{T_{K}}$. 
(b) Schematic susceptibility curve for the Kondo effect, showing
cross-over from Curie susceptibility at high temperatures to Pauli
susceptibility at temperatures below the Kondo temperature $T_{K}$.
(c) Specific heat curve for the Kondo effect. Since the total area is
the full spin entropy $R\ln  2$ and the width is of order $T_{K}$, the
height must be of order $\gamma\sim R\ln 2/T_{K}$. This sets the scale for the
zero temperature specific heat coefficient. 
}{chicvkondo}

Today, we understand this 
logarithmic correction as a renormalization\index{renormalization} of the Kondo coupling constant,
resulting  from
fact that as the temperature is lowered, more and more high frequency
quantum spin fluctuations become coherent, and these
strengthen the Kondo interaction. 
The effect is closely analogous to the growth of the strong-interaction
between quarks, and like quarks, the local moment in the Kondo
effect\index{Kondo effect}
is {\sl asymptotically free} at high energies. 
However, as you 
can see from the above
equation, once the temperature becomes of order
\[
T_{K}\sim D \exp  \left[-\frac{1}{2J\rho } \right]
\]
the correction becomes as large as the original perturbation, and at
lower temperatures, the Kondo interaction can no longer be treated
perturbatively.    In fact, non-perturbative methods 
tell us that  this interaction
scales to strong coupling at low energies, causing electrons
in the conduction sea to magnetically 
screen the local moment to form an inert {\sl Kondo
singlet}\index{Kondo singlet}
denoted by
\begin{equation}\label{}
\vert GS\rangle = \frac{1}{\sqrt{2}}\left( \vert \Uparrow \downarrow
\rangle - \vert \Downarrow \uparrow \rangle 
\right), 
\end{equation}
where the thick arrow refers to the spin state of the local moment
and the thin arrow refers to the spin state of a bound-electron at the
site of the local moment.  The key features of the impurity Kondo
effect\index{Kondo effect} are are: 
\begin{itemize}

\item The electron fluid surrounding the Kondo singlet forms a Fermi
liquid, with a Pauli susceptibility 
$\chi \sim 1/T_{K}$. 

\item The local moment is a kind of qubit which 
entangles with the conduction sea to form a singlet. 
As the temperature $T$ is raised, the entanglement entropy converts to
thermal entropy, 
given by the integral of the specific heat coefficient,
\[
S (T) = \int_{0}^{T}dT' \frac{C_{V} (T')}{T^{'}}.
\]
Since the total area under the curve, 
$S (T\rightarrow \infty)= R\ln 2$ per mole is the high temperature
spin entropy, and 
since the characteristic width  
is the Kondo temperature, it follows that the 
the characteristic zero temperature specific heat coefficient must be of order
the inverse Kondo temperature\index{Kondo temperature}:
$\gamma=\frac{C_{V}}{T} (T\rightarrow 0)\sim 
\frac{R\ln 2}{T_{K}}$. (See Fig. \ref{chicvkondo} b)

\item The only scale in the physics
is $T_{K}$. 
For example, the resistivity
created by magnetic scattering off the impurity 
has a universal temperature dependence 
\begin{equation}\label{}
\frac{R (T)}{R_{U}} = n_{i} \Phi  \left(\frac{T}{T_{K}}\right)
\end{equation}
where $n_{i}$ is the concentration of magnetic impurities, 
$\Phi  (x)$ is a universal function and $\rho_{U}$ is the unit of 
{\sl unitary resistance} (basically resistance with a scattering rate
of order the Fermi energy), 
\begin{equation}\label{}
R_{U}= \frac{2 n e^{2}}{\pi m \rho  }
\end{equation}
Experiment confirms that the resistivity in the Kondo
effect\index{Kondo effect}
can indeed be scaled onto a single curve that fits forms derived from
the Kondo model 
(see Fig \ref{kondores}).

\item The scattering off the Kondo singlet is resonantly confined to a narrow
region of order $T_{K}$, called the 
{\sl Kondo} or {\sl Abriksov-Suhl} resonance. 
\end{itemize}
\fight=0.5 \textwidth
\fg{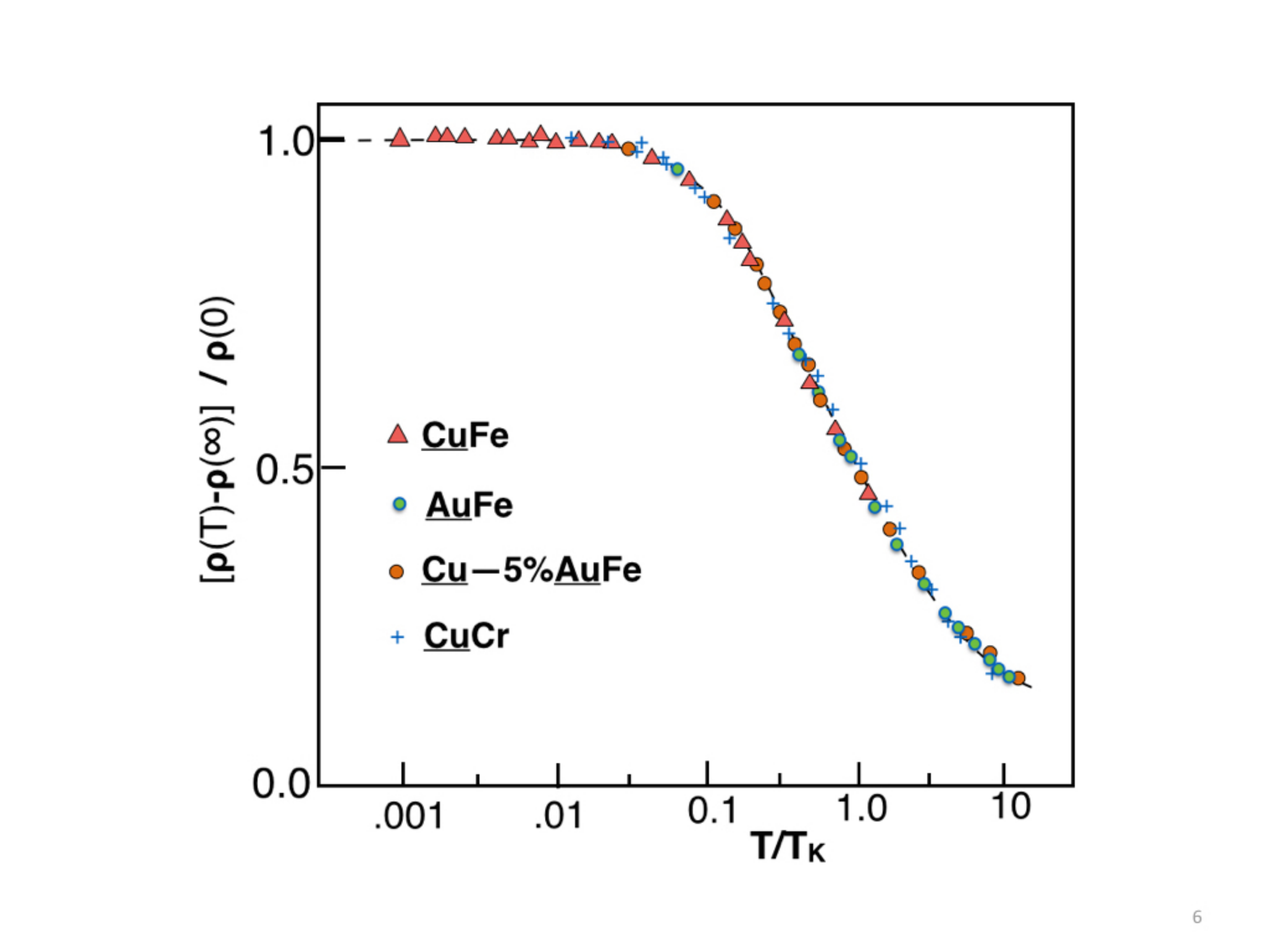}{
Temperature dependence of
resistivity associated with scattering from an impurity spin
from \cite{kondo_scaling,daybell}. The resistivity saturates at the unitarity
limit at low temperatures, due to the formation of the Kondo
resonance. Adapted from \cite{kondo_scaling}.
}{kondores}

\subsection{The Kondo lattice}\label{}\index{Kondo lattice}

In  heavy fermion material, containing a lattice of local moments,
the Kondo effect develops {\sl coherence}\index{coherence}.
In a single impurity, a Kondo singlet\index{Kondo singlet}
scatters electrons without conserving momentum, giving rise to a huge
build-up of resistivity at low temperatures. However, in a lattice,
with translational symmetry, 
this same elastic scattering now conserves momentum, and this leads to
coherent scattering off the Kondo singlets. 
In the simplest heavy fermion metals, this
leads to a dramatic reduction in the resistivity at temperatures below
the Kondo temperature. 

As a simple example, consider CeCu$_{6}$ a classic heavy
fermion metal.  Naively, 
CeCu$_{6}$ is just a
copper alloy, in which 
14\% of the copper atoms are replaced by cerium, yet this modest
replacement radically alters the metal. In this material, it actually
proves possible to follow the development of coherence\index{coherence} from the dilute
single ion Kondo limit, to the dense Kondo lattice\index{Kondo lattice}, by forming the
alloy La$_{1-x}$Ce$_{x}$Cu$_{6}$.
Lanthanum is iso-electronic to cerium, but has an empty f-shell, so
the limit $x\rightarrow 0$ corresponds to the dilute Kondo limit, and
in this limit the resistivity follows the classic Kondo
curve. However, as the concentration of cerium increases, the
resistivity curve starts to develop a coherence maximum, an in the
concentrated limit drops to zero with a characteristic $T^{2}$
dependence of a Landau Fermi liquid\index{Landau Fermi liquid} (see Fig. \ref{cohere}). 

CeCu$_{6} $ displays the following classic features of a heavy fermion
metal: 
\begin{itemize}
\item A Curie-Weiss susceptibility $\chi\sim (T+\theta )^{-1}$ at high temperatures.

\item  A paramagnetic spin susceptibility $\chi \sim \hbox{cons}$
at low temperatures. 

\item A dramatically enhanced  linear specific heat  $C_{V}= \gamma T$ at low
temperatures,  where in CeCu$_{6}$
$\gamma\sim 1000 $mJ/mol/K$^{2}$
is about $1000$ times larger than in copper. 

\item A quadratic temperature dependence of the low temperature
resistivity $\rho = \rho _{o}+ A T^{2}$

\end{itemize}

In a Landau Fermi liquid\cite{landaufl}, 
the magnetic susceptibility $\chi $ and the linear specific heat 
coefficient $\gamma = C_{V}/T\vert_{T\rightarrow 0}$ are given by 
\begin{eqnarray}\label{}
\chi &=& (\mu_{B})^{2}\frac{N^* (0)}{1 + F_{o}^{a}}\cr
\gamma &=& \frac{\pi^{2}k_{B}^{2}}{3} N^{*} (0)
\end{eqnarray}
where $N^{*} (0)= \frac{m^{*}}{m}N (0)$ is the renormalized
density of states and $F_{0}^{a}$ is the spin-dependent part 
of the s-wave interaction between quasiparticles.  One of the
consequences of Fermi liquid theory, is that the density of states
factors out of the Sommerfeld or Wilson
ratio between the susceptibilty and linear specific heat coefficient,
\begin{equation}\label{}
W = \frac{\chi }{\gamma} = \left( \frac{\mu_B}{2\pi k_{B}}\right)^{2}\frac{1}{1+F_{0}^{a}}.
\end{equation}
In heavy fermion metals, 
this ratio remains approximately fixed across several decades of 
variation in $\chi$ and $\gamma$. This allows us 
to understand heavy fermion metals 
as a lattice version of the Kondo effect
gives rise to a renormalized density of states $N^{*} (0)\sim
\frac{1}{T_{K}}$. 

The discovery of heavy electron compounds in the 1970s 
led Mott\cite{Mott:1974ui} and Doniach\cite{DONIACH:1977um} 
to propose that heavy electron systems should be modeled as a
``Kondo-lattice'',
where a dense array of local moments  interact with the conduction
sea via an antiferromagnetic interaction $J$.
In such a lattice, the local moments polarize the conduction sea, and
the resulting Friedel oscillations in the magnetization give rise to
an antiferromagnetic 
RKKY (Rudermann Kittel Kasuya Yosida) magnetic
interaction\cite{rkky,rkky2,rkky3}\index{RKKY interaction} that
tends to order the local moments. Mott and
Doniach realized that this interaction must compete with the Kondo
effect. 

The simplest Kondo lattice Hamiltonian\cite{Kasuya:1956da}\index{Kondo
lattice}is
\begin{equation}\label{}
H=\sum_{{\bk}\sigma }\epsilon_{{\bk}}c\dg _{{\bk}\si
}c_{{\bk}\si}
+ J\sum_{j} \vec{S}_{j}\cdot c\dg _{j\alpha
}
\vec{\sigma }
_{\alpha \beta }c_{j\beta
},\end{equation}
where
\begin{equation}\label{}
c\dg _{j\alpha }= \frac{1}{\sqrt{\Nsites}}\sum_{\bk }
c\dg _{\bk \alpha }e^{-i \bk \cdot \bR_{j}}
\end{equation}
creates an electron at site $j$. 
Mott and Doniach\cite{Mott:1974ui,DONIACH:1977um} pointed out that there are two
 energy
scales in the  Kondo lattice\index{Kondo lattice}: the Kondo
temperature $T_{K} \sim D e^{-1 / (2 J \rho) }$ and the RKKY scale 
$E_{RKKY}=E_{RKKY}= J^{2}\rho $\index{RKKY interaction}. 
\fight=0.6\textwidth \fg{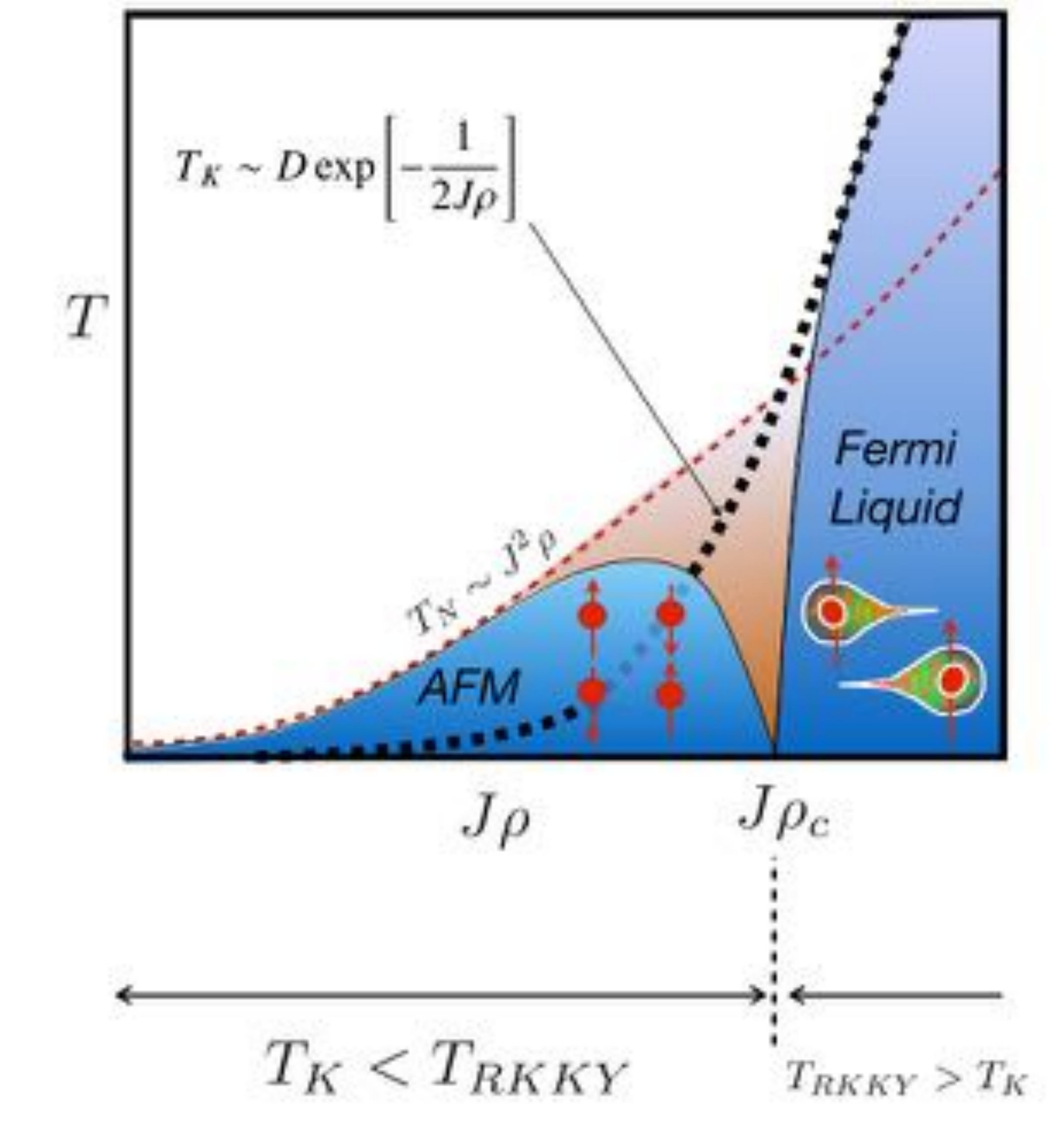}{Doniach phase
diagram\index{Doniach phase diagram}
for the Kondo lattice, illustrating the antiferromagnetic
regime
 and the heavy fermion regime, for  $T_{K}<T_{RKKY}$ and $T_{K}>
T_{RKKY}$ respectively.
The effective 
Fermi temperature of the heavy Fermi liquid is indicated as a solid
line. Experimental evidence suggests that in many
heavy
fermion materials this scale
drops to zero at the antiferromagnetic quantum critical point. 
}{fig16}
For small  $J\rho  $, $E_{RKKY}>> T_{K}$ leading to an antiferromagnetic
ground-state, but when $J\rho $ is large, 
$T_{K}>>E_{RKKY}$, stabilizing a ground-state in which
every site in the lattice resonantly scatters electrons. 
Based on a 
a simplified one-dimensional ``Kondo necklace'' model\cite{doniachjullien},
Doniach conjectured\cite{DONIACH:1977um} that the
transition between the antiferromagnet and the dense Kondo
ground state is a continuous quantum phase transition.
Experiment confirms this conjecture, 
and today we have several examples of
such quantum critical points, including 
CeCu$_{6}$ doped with gold to form
CeCu$_{6-x}$Au$_{x}$ and CeRhIn$_{5}$ under pressure\cite{tusonpark,revqcp2,revqcp3}. 
In the fully developed Kondo lattice ground state
Bloch's theorem insures
that the resonant elastic scattering at each site will generate 
a renormalized f- band, of width $\sim T_{K}$. 
In contrast with the impurity Kondo effect, here
elastic scattering at each site
acts coherently. 
For this reason, as the heavy electron metal develops at low
temperatures, its  resistivity drops towards zero (see Fig. \ref{cohere}b). 

In a Kondo lattice, spin entanglement is occurring on a truly
macroscopic scale, but this entanglement need not necessarily 
lead to a Fermi liquid.
Experimentally, many other possibilities are possible. Here are
some examples, 
\begin{itemize}
\item 
Ce$_{3}$Bi$_{4}$Pt$_{3}$, a Kondo insulator\index{Kondo insulator} in which the formation of 
Kondo singlets with the Ce moments drives the development of a small 
insulating gap at low temperatures and 
\item 
CeRhIn$_{5}$, an antiferromagnet on the brink of forming a Kondo
lattice, which under pressure becomes a heavy fermion
superconductor with $T_{c}$=2K.

\item UBe$_{13}$ a heavy fermion 
superconductor which transitions directly from an
incoherent metal with resistivity 200$\mu\Omega$cm, into a 
superconducting state. 

\end{itemize}
Each of these materials has qualitatively the same 
same high temperature Curie Weiss magnetism and
the same Kondo resistivity at high temperatures, due to incoherent
scattering off the local moments.
However at low
temperatures the scattering off the magnetic Ce ions 
becomes coherent and new properties
develop.    

\fight=0.7\textwidth
\fg{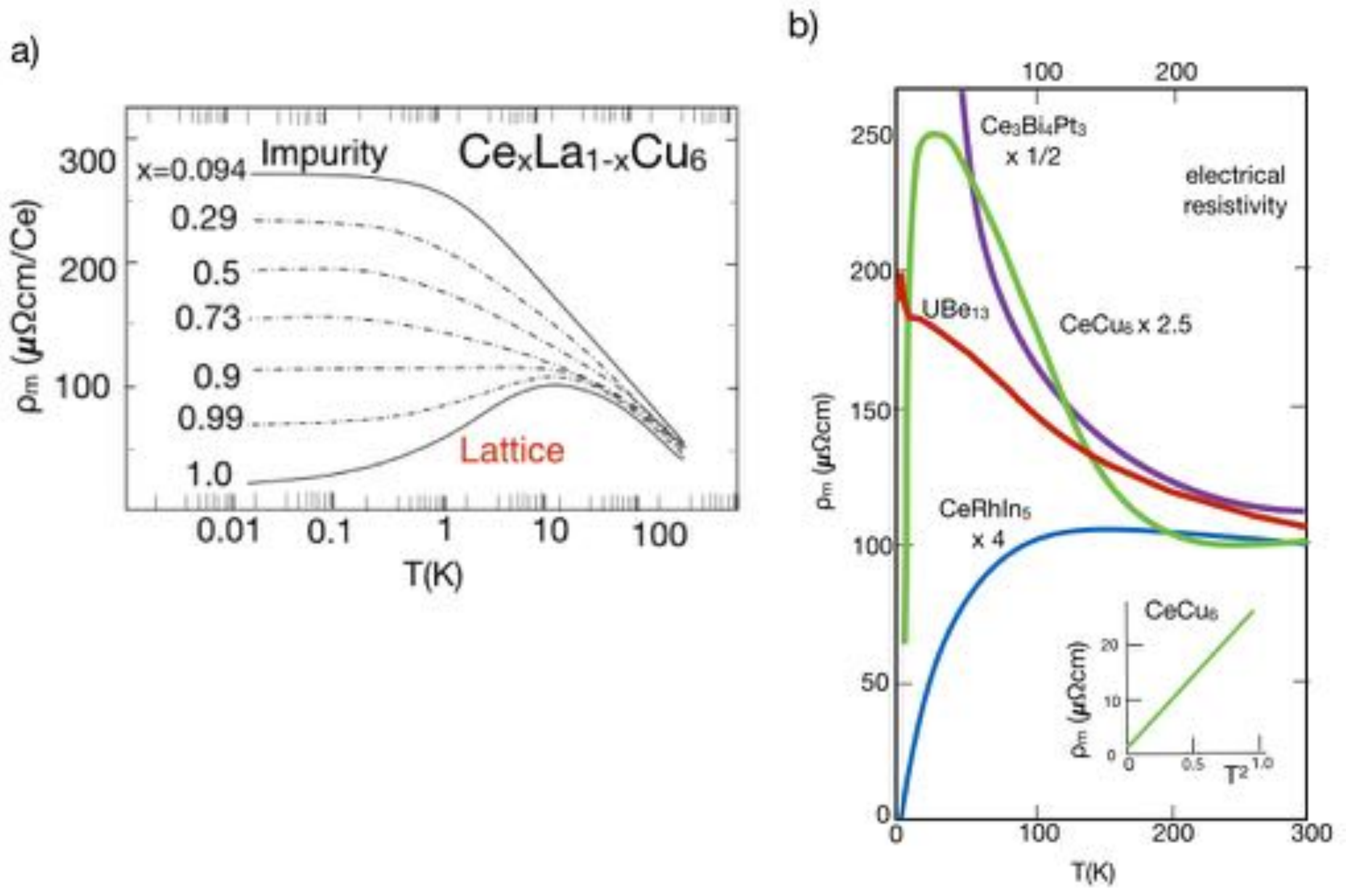}{
(a) Resistivity of Ce$_{x}$La$_{1-x}$Cu$_{6}$. Dilute
Ce atoms in LaCu$_{6}$ exhibit a classic ``Kondo''
resistivity, but as the Ce concentration becomes dense,
elastic scattering off each Ce atom leads to the
development of a coherent heavy fermion metal. (b) 
Resistivities of four heavy fermion materials showing the development
of coherence\index{coherence}. A variety of antiferromagnetic
magnetic, Fermi liquid, superconducting and insulating states are
formed (see text).
}{cohere}


\section{Kondo insulators: the simplest heavy fermions
}\label{}\index{Kondo insulator}

In many ways, the Kondo insulator is the simplest ground-state of the
Kondo lattice.   The first Kondo insulator (KI), \SMB was discovered
almost fifty
years ago~\cite{smb6} and today there are several 
known examples including Ce$_{3}$Bi$_{4}$Pt$_{3}$. 
At room temperature, these KIs
are metals containing a dense array of
magnetic moments, yet on cooling they 
develop a narrow gap
due 
the formation of {\sl Kondo singlets} which screen the 
local moments~\cite{revfiskaeppli,Fisk96,revki1,revki2}.
We can gain a lot of insight by
examining the strong coupling limit in which the dispersion of the
conduction sea is much smaller than the Kondo coupling $J$. 
Consider a simple tight-binding Kondo lattice 
\begin{equation}\label{}
H=-t\sum_{(i,j)\sigma } (c\dg_{i\sigma }c_{j\sigma }+{\rm H.c})
+ {J}\sum_{j, \alpha \beta } \vec{\sigma }_{j}
\cdot\vec S_{j}, \qquad \qquad \vec{ \sigma }_{j}\equiv ( c \dg _{j\beta }\vec{\sigma }_{\beta \alpha }c _{j\alpha }
)
\end{equation}
in which $t/J<<1$ is a small
parameter.  In this limit,  the inter-site hopping is a perturbation 
to the on-site Kondo interaction, 
\begin{equation}\label{}
H\stackrel{t/J\rightarrow 0}\longrightarrow
{J}\sum_{j, \alpha \beta } 
\vec{\sigma }_{j} 
\cdot\vec S_{j}+O (t),\
\end{equation}
and the corresponding ground-state 
corresponds to the formation of a spin 
singlet at each site, denoted by the wavefunction
\begin{equation}\label{}
\vert KI \rangle=\prod_j \frac{1}{\sqrt{2}}\biggl( \Uparrow_{j} \downarrow_{j}-\Downarrow_{j}
\uparrow_{j} \biggr) 
\end{equation}
where the double and single arrows denote the localized moment and
conduction electron respectively.

Each singlet
has a ground-state energy $E = - \frac{3}{2}J$ per site and a singlet-triplet
spin gap of magnitude $\Delta E = 2J$. Moreover, if 
we either remove an electron from site i, 
we break a Kondo singlet and 
create an unpaired spin with excited energy  $\frac{3}{2}J$, 
\begin{equation}\label{}
|\hbox{qp}^{+},i \uparrow \rangle  =  \Uparrow_{i}\prod_{j\neq i} \frac{1}{\sqrt{2}}\biggl( \Uparrow_{j} \downarrow_{j}-\Downarrow_{j}
\uparrow_{j} \biggr) = \sqrt{2} c_{i\downarrow}\vert KI\rangle,
\end{equation}
as illustrated in Fig \ref{kondoinsfig2}(a).
Similarly, if we add an electron, we create an electron quasiparticle,
corresponding to an unpaired local moment and a doubly occupied
conduction electron orbital
\begin{eqnarray}\label{l}
|\hbox{qp}^{-},i \uparrow \rangle  =
 \Uparrow_{i}
\biggl(\uparrow_{i}\downarrow_{i}\biggr)
\prod_{j\neq i} \frac{1}{\sqrt{2}}\biggl( \Uparrow_{j}
\downarrow_{j}-\Downarrow_{j}\uparrow_{j} \biggr) = \sqrt{2}c\dg_{j\uparrow}\vert KL\rangle,
\end{eqnarray}
as illustrated in Fig \ref{kondoinsfig2}(b).

\fight=0.7 \textwidth
\fgb{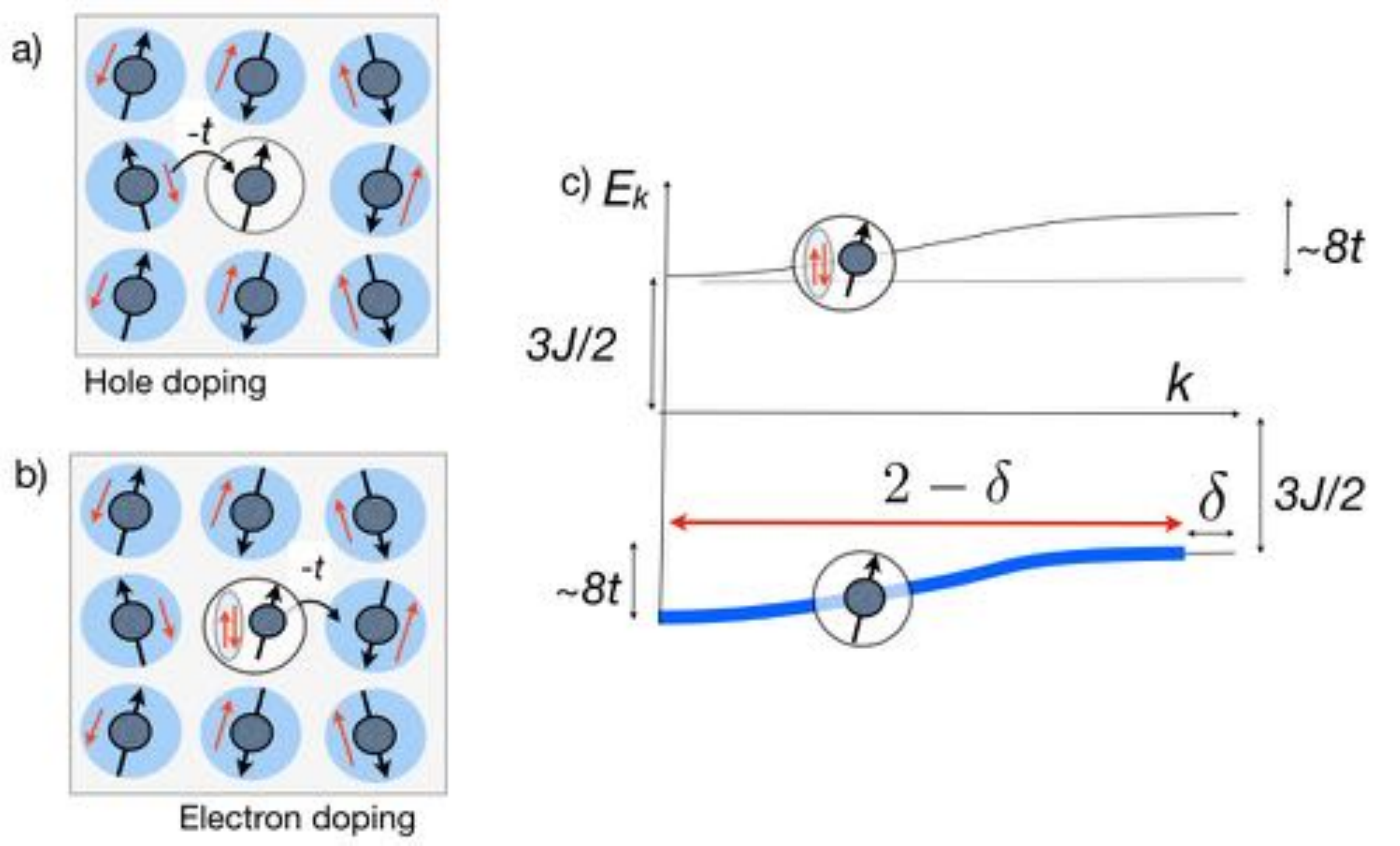}{Showing (a) hole and (b) electron
doping of strong coupling Kondo insulator\index{Kondo insulator}. (c) Dispersion of strong
coupling Kondo insulator. A small amount of hold doping $\delta $
gives rise to a ``large'' Fermi surface containing $2-\delta $
heavy electrons. 
}{kondoinsfig2}
\fight=0.3 \textwidth

If we now reintroduce the hopping $-t$ between sites, then these
quasiparticle excitations become mobile, as illustrated in
Fig. \ref{kondoinsfig2}(a) and (b). From the explicit form of the
states, we find that the nearest neighbor hopping matrix elements
are  $\langle  {\hbox{qp}^{\pm},i \sigma } \vert H \vert \hbox{qp}^{\pm},j\sigma \rangle
= \pm\frac{t}{2}, 
$,
giving quasiparticle energies
\begin{equation}\label{}
E_{\hbox{qp$^{\pm}$}} (\bk ) =\pm t (c_{x}+c_{y}+c_{z})+ \frac{3}{2}J.
\end{equation}

To transform from the quasiparticle, to the electron basis, we need to
reverse the sign of the hole (qp$^{+}$) dispersion to obtain the
valence band dispersion, so that the band energies
predicted by the strong coupling limit 
of the Kondo lattice are 
\begin{equation}\label{}
E^{\pm }_{\bk }= -t (c_{x}+c_{y}+c_{z})\pm  \frac{3}{2}J,
\end{equation}
separated by an energy $3J$ as shown in Fig. \ref{kondoinsfig2}(c).
Note that these are ``hard core'' fermions that can not occupy the
same lattice site simultaneously. 

In this way, the half-filled strong coupling Kondo lattice forms an
insulator with a charge gap of size $3J$ and a spin gap of size
$2J$. Notice finally that if we dope the insulator with an amount
$\delta $ of holes, we form a band of heavy fermions. In this way,
Kondo insulators\index{Kondo insulator} can be considered the parent states of heavy electron
materials.  However, we'd like to examine the physics of a Kondo
lattice at weak coupling, and to do this requires a different
approach. 

\section{Large $N$ Expansion for  the Kondo Lattice}\index{large N expansion}

\subsection{Philosophy and Formulation}\label{}

One of the great difficulties with the Kondo lattice, is that there is
no natural small parameter to carry out an approximate treatment.  One
way around this difficulty, is to use a large $N$ expansion, in which
we extend the number of spin components of the electrons from $2$ to
$N$.  Historically, 
Anderson\cite{andersonlargeN} pointed out that the large spin-orbit
coupling in heavy fermion compounds generates (if we ignore crystal
fields) 
a large spin degeneracy $N=2j+1$, furnishing 
a small parameter $1/N$ for a controlled expansion about 
the limit $N\rightarrow \infty $.   One of the observations arising
from Anderson's idea \cite{me1983,read83_1} is that the RKKY
interaction\index{RKKY interaction} becomes negligible (of order $O (1/N^{2})$) in this
limit and the Kondo lattice ground-state
becomes stable. This observation opened the way to path integral
mean-field treatments of the Kondo lattice \cite{read83_1,slaveb,read83_v2,auerbach,millislee,long}

The basic idea of the large $N$ limit\index{large N expansion} is to examine a limit 
where every term in the Hamiltonian grows extensively with $N$.
In the path integral for the partition function,
the corresponding action then grows extensively with $N$, so that
\begin{equation}\label{}
Z = \int {\cal D}[\psi ] e^{-NS}= \int {\cal D}{\psi }\exp
 \left[
 {- \frac{S}{1/N}}
\right]\equiv 
\int {\cal D}[\psi ]\exp \left[ {- \frac{S}{\hbar_{eff}}}\right].
\end{equation}
Here  $\frac{1}{N} \sim \hbar_{eff}$
behaves as an effective Planck's constant for the theory,
focusing the path integral into a non-trivial ``semi-classical'' or
``mean field'' solution as 
$\hbar_{eff}\rightarrow 0$. 
As $N\rightarrow \infty $, the
quantum  fluctuations of intensive variables $\hat a$, such as the
electron density per spin, become smaller and smaller, scaling as $\langle
\delta a^{2}\rangle /\langle a^{2}\rangle \sim 1/N
$, causing the path integral to focus around a non-trivial mean-field trajectory.  In this way, 
one can obtain new results by expanding around the 
solvable large $N $ limit\index{large N expansion} in powers of $\frac{1}{N}$. 
(Fig. \ref{fig15x}). 
\fight=0.4\textwidth
\fg{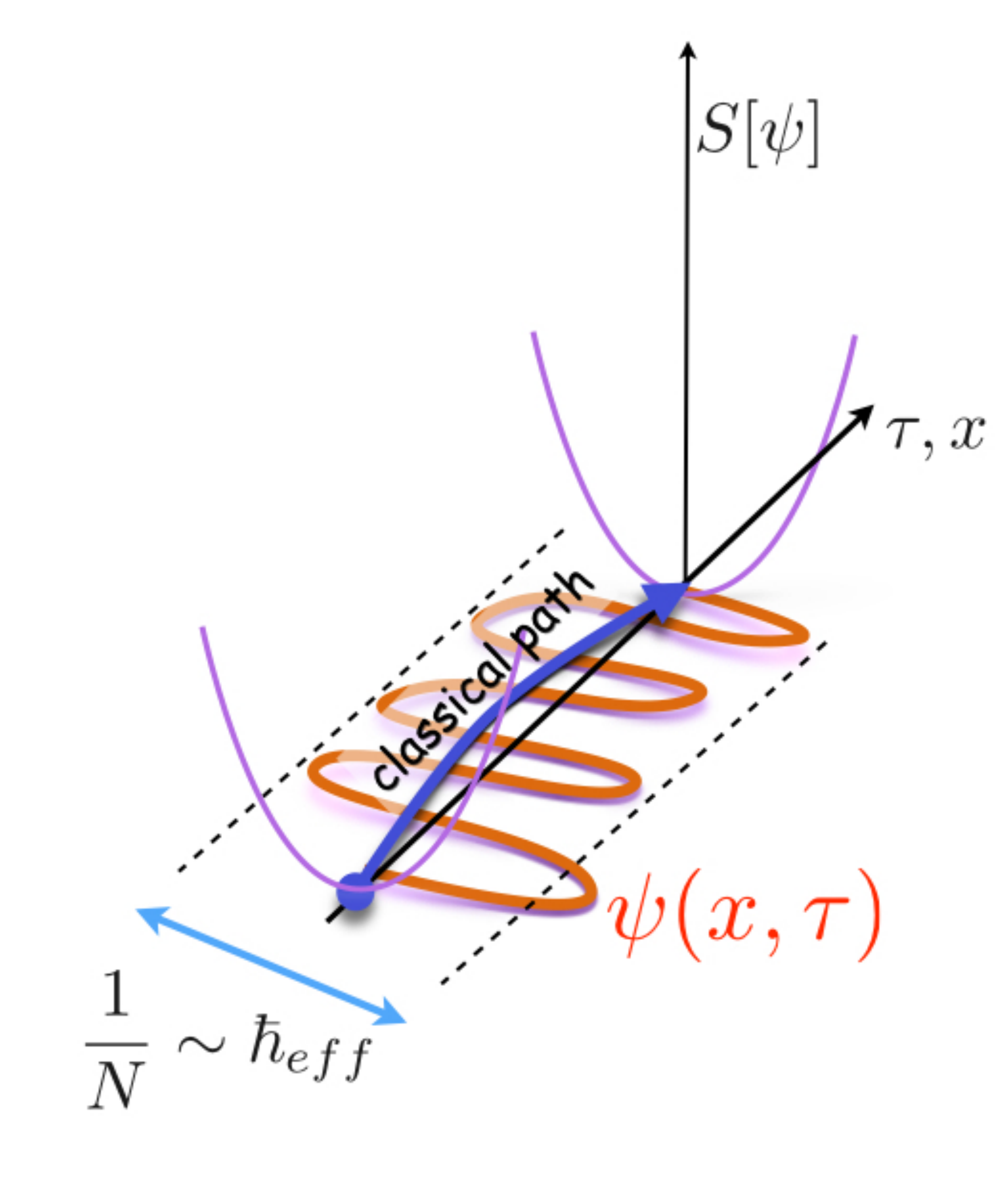}{Illustration of the 
convergence of a quantum path integral about a semi-classical
trajectory in the large $N$ limit. }{fig15x}

We will use a simplified Kondo lattice model introduced
by Read and Newns\cite{read83_1}, 
in which all electrons have a spin degeneracy
$N=2j+1$, 
\begin{equation}\label{}
H=\sum_{{\bk}\alpha }\epsilon_{{\bk}}c\dg _{{\bk}\alpha
}c_{{\bk}\alpha}
+ \frac{J}{N}\sum_{j, \alpha \beta } 
c \dg _{j\beta }c _{j\alpha }S_{\alpha \beta } (j).
\end{equation}
where $c\dg _{j\alpha }= \frac{1}{\sqrt{\Nsites}}\sum_{{\bk}}c\dg _{{\bk}\alpha
}e^{-i{\bk}\cdot \vec{R}_{j}}
$
creates an electron localized at site $j$  
and the spin of the local moment at position  $\bR_j$
is represented by pseudo-fermions
\begin{equation}\label{}
S_{\alpha \beta } (j)=f\dg _{j\alpha } f_{j\beta }- \frac{n_{f} (j)}{N}\delta_{\alpha \beta }.
\end{equation}
This representation requires that we set a value for the conserved $f$
occupancy $n_{f} (j)=Q$ at each site. 
This interaction can be rewritten in a factorized 
form
\sboxit{\begin{equation}\label{}
H=\sum_{{\bk}\alpha }\epsilon_{{\bk}}c\dg _{{\bk}\alpha
}c_{{\bk}\alpha}
- \frac{J}{N}\sum_{j, \alpha \beta } 
:\left(c \dg _{j\beta }f_{j\beta } \right)
\left(f\dg _{j\alpha } c_{j\alpha } \right):
\end{equation}
\rightline{\bf Read Newns model for the Kondo lattice}}
where the potential scattering terms resulting from the rearrangement
of the f-operators have been absorbed into a shift of the chemical
potential. 
Note that:
\begin{itemize}
\item the model has a global $SU (N)$ symmetry associated with the
conserved magnetization. 

\item the Read Newns (RN) model is a lattice version of the 
Coqblin-Schrieffer Hamiltonian\cite{coqblin}
introduced to describe the Kondo interaction in
strongly spin-orbit coupled rare-earth ions. 
While the Coqblin-Schrieffer interaction\index{Coqblin Schrieffer interaction} is correct at each site, 
the assumption that the  $SU (N)$
spin is conserved by electron hopping is an oversimplification. 
(This is a price one pays for a solvable model.)

\item  in this factorized form, the antiferromagnetic Kondo interaction is
``attractive''.
\item the coupling constant has been scaled to vary as
$J/N$, to ensure that the interaction grows extensively with $N$. 
The interaction 
involves a product of two terms that scale as $O (N)$,
so that $J/N\times
O (N^{2}) \sim O (N)$.

\item 
the RN model also has a {\bf local gauge
invariance}: the absence of f-charge
fluctuations allows us to change the
phase of the f-electrons {\sl independently } at each site
\begin{equation}\label{}
f_{j\sigma }\rightarrow e^{i\phi_{j}}f_{j\sigma }.
\end{equation}

\end{itemize}
A tricky issue concerns the 
value we give to the conserved 
charge $n_{f}=Q$. In
the physical models of interest,
$n_{f}=1$ at each site, so one might be inclined to explicitly maintain this
condition.  However, the large $N$
expansion requires that  the action is extensive in  $N$, and
this forces us to consider
more general classes of
solution where $Q$ scales
with $N$ so that 
the filling factor
$q= Q/ N$ is finite as $N\rightarrow \infty $. 
Thus if we're interested 
in a Kramer's doublet Kondo model, we  take the
half-filled case $q=1/2$,
$Q=N/2$, but if we want to understand a 
$j=7/2$ Yb$^{3+}$ atom without crystal fields, then in the physical
system $N=2j+1=8$, and we should fix $q=Q/N=1/8$. 

The partition function for the Kondo lattice is then
\begin{equation}\label{}
Z = {\rm Tr} \left[e^{-\beta H}\prod_{j}\delta (\hat n_{f} (j)-Q)\right]
\end{equation}
where $\delta (\hat n_{f} (j)-Q) $ projects out the states
with $n_{f} (j) =Q$ at site $j$.  By re-writing the delta function as
a Fourier transform, the partition function can be
can be rewritten as a path-integral, 
\begin{equation}\label{}
Z = \int {\cal D}[ \psi\dg  , \psi , \lambda] \exp \left[-
\int_{0}^{\beta } d\tau
\overbrace {
\left( 
\psi\dg   \partial_{\tau }\psi + H[\bar \psi ,\psi
,\lambda]\right)}^{L[ \psi\dg  ,\psi , \lambda]}\right]
\end{equation}
where
$\psi\dg  \equiv (\{c\dg \},\{ f\dg  \})$ schematically 
represent the conduction and f-electron fields,
\begin{equation}\label{}
H[\lambda] =\sum_{{\bk}\alpha }\epsilon_{{\bk}}c\dg _{{\bk}\alpha
}c_{{\bk}\alpha}
- \frac{J}{N}\sum_{j, \alpha \beta } 
:\left(c \dg _{j\beta }f_{j\beta } \right)
\left(f\dg _{j\alpha } c_{j\alpha } \right): + \sum_{j}\lambda_{j} (n_{fj} -Q).
\end{equation}
The field $\lambda_{j}$ is a fluctuating Lagrange multiplier
that enforces the constraint $n_{j}=Q$ at each site.

Next we carry out a Hubbard Stratonovich
transformation on the interaction, 
\begin{equation}\label{}
-\frac{J}{N}\sum_{ \alpha \beta }\left(c \dg _{j\beta }f_{j\beta } \right)
\left(f\dg _{j\alpha } c_{j\alpha } \right)
\rightarrow 
\sum_{\alpha } \left[
\bar V_{j}\left(c \dg _{j\alpha  }f_{j\alpha  } \right)
+
\left( f\dg _{j\alpha } c_{j\alpha } \right)V_{j} \right]
 +N\frac{\bar V_{j}V_{j}}{J}.
\end{equation}
In the original Kondo model, 
we started out with an interaction between electrons and spins. Now,
by carrying out the
Hubbard Stratonovich transformation, 
we have formulated the interaction as the exchange of a charged
boson
\begin{eqnarray}\label{kondoboson}
\parbox{25mm}{\bfyn(25,20)
\fmfset{arrow_len}{3mm}
\fmfleftn{i}{11}\fmfrightn{o}{11}
\fmf{fermion,right=0.0}{v1,i10}
\fmf{dashes_arrow}{i2,v1}
\fmf{dashes_arrow}{v1,o10}
\fmf{fermion}{o2,v1}
\fmfv{decor.shape=circle,decor.filled=30,decor.size=12thick,label=
$\frac{J}{N}$,label.dist=0}{v1}
\efyn}
&\equiv&\ \ 
\parbox{45mm}{\bfyn(45,20)
\fmfset{arrow_len}{3mm}
\fmfleftn{i}{11}\fmfrightn{o}{11}
\fmf{fermion,right=0.0}{v1,i10}
\fmf{dashes_arrow}{i2,v1}
\fmf{photon,tension=1,label=$\frac{J}{N}\delta (\tau-\tau')$ }{v2,v1}
\fmf{dashes_arrow}{v2,o10}
\fmf{fermion}{o2,v2}
\fmffreeze
\fmfv{label=$c\dg_{\beta }f_{\beta }\ \ $
}{v1}
\fmfv{label=$\ \ f\dg _{\alpha }c_{\alpha}$
}{v2}
\efyn}
\cr
- \frac{J}{N}
\sum_{\bk ,\bk ',\alpha ,\beta }
(c\dg_{\beta }f_{\beta })
(f\dg _{\alpha}c_{\alpha})&&
\end{eqnarray}
where the solid lines represent the conduction electron propagators,
and the dashed lines represent the f-electron operators. 
Notice how 
the bare amplitude associated with the exchange boson is
frequency independent, i.e the interaction is instantaneous. 
Physically, we may
interpret this exchange process as due an intermediate valence
fluctuation.

The path integral now involves an additional integration over
the hybridization fields $V$ and $\bar V$, 
\sboxit{
\begin{eqnarray}\label{matrixyoulove}
Z&=& \int {\mathcal{D}}[\bar V,V,\lambda ] 
\int{\cal D}[\psi \dg,\psi] 
\exp \biggl[
-\overbrace {
\int_{0}^{\beta } ( \psi \dg \partial_{\tau }\psi  +
H[\bar V, V,\lambda ]) 
}^{S[\bar V,V,\lambda,\  \psi\dg ,\psi]}\biggr]\cr\cr
H[\bar V, V,\lambda ]&=&\sum_{{\bk}}\epsilon_{{\bk}}c\dg _{{\bk}\si
}c_{{\bk}\si}
+ \sum_{j} 
\left [
\bar V_{j}\left(c \dg _{j\sigma  }f_{j\sigma  } \right)
+ \left( f\dg _{j\sigma } c_{j\sigma } \right)V_{j}
+\lambda _{j} ( n_{fj}-Q) 
+ N \frac{\bar V_{j}V_{j}}{J}\right],
\end{eqnarray}
\rightline {\bf Read Newns path integral for the Kondo lattice.}}
where we have suppressed summation signs for repeated
spin indices (summation convention).


The RN path integral allows us
to develop a mean-field description of the many body Kondo scattering
processes that captures the physics and is asymptotically exact as
$N\rightarrow \infty $. In this approach, 
the condensation of the hybridization field 
describes the formation of bound-states between spins
and electrons that can not be dealt with in perturbation theory.  
Bound-states induce long range temporal correlations in scattering:
once the hybridization condenses, 
the interaction
lines break-up into independent anomalous scattering events, denoted
by
\[
\parbox{45mm}{\bfyn(45,20)
\fmfset{arrow_len}{3mm}
\fmfleftn{i}{11}\fmfrightn{o}{11}
\fmf{fermion}{i10,v1}
\fmf{dashes_arrow}{v1,i2}
\fmf{photon,tension=0.8,label.dist=4.,label.side=left,label= $\langle \delta \bar V (1)\delta V (2)\rangle$}{v1,v2}
\fmf{dashes_arrow}{o10,v2}
\fmf{fermion}{v2,o2}
\fmffreeze
\fmf{phantom_arrow,tension=2}{v2,v1}
\efyn}
\rightarrow 
\parbox{45mm}{\bfyn(45,20)
\fmfleftn{i}{11}\fmfrightn{o}{11}
\fmf{fermion,tension=0.8,right=0.0}{i10,v1}
\fmf{dashes_arrow,tension=1.2}{v1,i2}
\fmf{phantom,tension=0.6}{v1,v2}
\fmfdot{v1}
\fmfv{decor.shape=circle,decor.filled=full,fore=red,decor.size=2thick,label=$\bar V (1)$,label.angle=0}{v1}
\fmfdot{v2}
\fmfv{decor.shape=circle,decor.filled=full,fore=red,decor.size=2thick,label=$\bar V (2)$,label.angle=180}{v2}
\fmf{dashes_arrow,tension=1.2}{o10,v2}
\fmf{fermion,tension=0.8}{v2,o2}
\fmffreeze
\efyn}
\]
The hybridization $V$ in the RN action 
carries the local $U (1)$ gauge charge
of the f-electrons, giving rise to an important local gauge invariance:
\sboxit{
\begin{eqnarray}\label{ReadNewnsGt}
f_{j\sigma }\rightarrow e^{i\phi_{j}}f_{j\sigma }, \qquad
V_{j}\rightarrow e^{i\phi_{j}}V_{j}, \qquad \qquad 
\lambda_{j}\rightarrow \lambda_{j}  -i\dot\phi_j (\tau ).
\end{eqnarray}
\rightline {\bf Read Newns gauge transformation.}
}
This invariance can be used 
to choose a gauge
in which $V_{j}$ is real, by absorbing 
the phase of the hybridization $V_{j}= |V_{j}|e^{i \phi_{j}}$
into the f-electron. 
In the radial gauge, 
\sboxit{
\begin{eqnarray}\label{matrixyoulove2}
Z&=& \int {\mathcal{D}}[ |V|,\lambda ] 
\int{\cal D}[\psi \dg,\psi] 
\exp \biggl[
-\overbrace {
\int_{0}^{\beta } ( \psi \dg \partial_{\tau }\psi  +
H[| V|,\lambda ]) 
}^{S[| V|\lambda,\  \psi\dg ,\psi]}\biggr]\cr\cr
H[|V|,\lambda ]&=&\sum_{{\bk}}\epsilon_{{\bk}}c\dg _{{\bk}\si
}c_{{\bk}\si}
+ \sum_{j} 
\left [
|V_{j}|
\left(c \dg _{j\sigma  }f_{j\sigma  } 
+ f\dg _{j\sigma } c_{j\sigma } \right)
+\lambda _{j} ( n_{fj}-Q) 
+ N \frac{|V_{j}|^{2}}{J}\right],
\end{eqnarray}
\rightline {\bf Read Newns path integral: ``radial gauge''.}}
Subsequently, when we use the radial
gauge, we will drop the modulii sign. 
The interesting feature about this Hamiltonian, is that with the real
hybridization, the conduction and f-electrons now transform under a single
global $U (1)$ gauge transformation, i.e the f-electrons have become
{\sl charged}. 

\subsection{Mean-Field Theory}\label{}

The interior fermion integral in the path integral (\ref{matrixyoulove2})
defines an effective action $S_{E}[V,\lambda]$ by the relation
\begin{eqnarray}\label{effact2}
Z_{E} = \exp 
\left[{-N S_{E}[V,\lambda]} \right]
\equiv 
\int {\cal D}[\psi \dg , \psi ] 
\exp \left[{-S[V,\lambda ,\ \psi \dg,\psi ]}  \right]
\ 
,
\end{eqnarray}
The extensive growth of the effective action with $N$
means that at large $N$, 
the integration in (\ref{matrixyoulove})
is dominated by its stationary points, allowing us to dispense with
the integrals over $V$ and $\lambda$. 
\begin{equation}\label{stationaryrules}
Z = \int {\cal D}[\lambda, V] \exp 
\left[{-NS_{E}[V,\lambda ]}
 \right]\approx 
\left. \phantom{\int_{0}^{\beta }}\hskip -0.4cm\exp 
\left[{-NS_{E}[V, \lambda]} \right]
\right\vert_{\hbox{Saddle
Point}} 
\end{equation}
In practice, we seek uniform, static solutions, 
$
V_{j} (\tau )=  V,\  \lambda_{j} (\tau ) =\lambda $. In this
case the saddle point partition function $Z_{E} 
={\rm  Tr}e^{-\beta H_{MFT}}
$ 
is simply the partition function of the static mean-field Hamiltonian
\begin{eqnarray}\label{l}
H_{MFT}= \sum_{{\bk}\sigma }
\left(c\dg _{{\bk}\sigma },f\dg
_{{\bk}\sigma } \right)
\overbrace {
\pmat{\epsilon _{{\bk}}& V\cr
\bar V&\lambda }
}^{\underline{h} (\bk )}
\pmat{c_{{\bk}\sigma }\cr f_{{\bk}\sigma }}
&+&N{\cal N}_{s}\left(\frac{|V|^{2}}{J}- \lambda q\right)
\\  \nonumber\cr
= \sum_{{\bk}\sigma } \psi \dg_{\bk \sigma }\ 
{\underline{h} (\bk )}\ \psi_{\bk \sigma }
&+&N{\cal N}_{s}\left(\frac{|V|^{2}}{J}- \lambda
q\right).\end{eqnarray}
Here,  $f\dg _{\bk \sigma }= \frac{1}{\sqrt{{\cal N}_{s}}}\sum_{j}f\dg _{j\sigma }
e^{i \bk \cdot \bR_{j}}$
is the Fourier transform of the $f-$electron field and 
we have introduced the two component notation 
\begin{equation}\label{Ch16:twocomponent}
\psi_{\bk \sigma } 
= \pmat{c_{{\bk}\sigma }\cr f_{{\bk}\sigma }}, 
\qquad 
\psi\dg _{\bk \sigma } =
\left(c\dg _{{\bk}\sigma },f\dg, 
_{{\bk}\sigma } \right), \qquad 
\underline{h} (\bk )= 
\pmat{\epsilon _{{\bk}}& V\cr
\bar V&\lambda }.
\end{equation}

\fight=4 truein
\fg{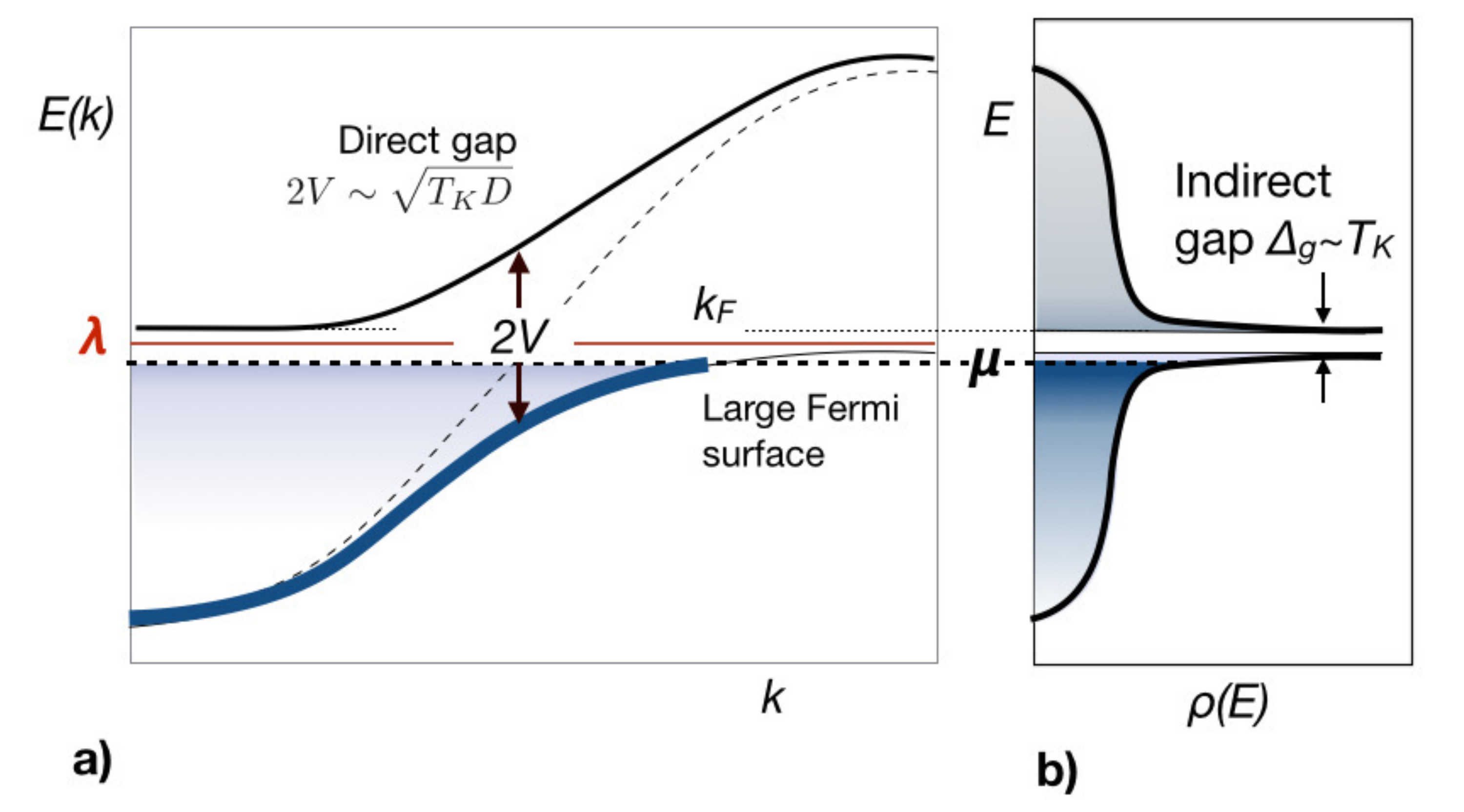}{(a) Dispersion for the Kondo lattice
mean field theory. (b) Renormalized density of states, showing ``hybridization gap''
($\Delta _{g}$).
}{fig20}

\fight=2.5 truein

We should think about $H_{MFT}$ as a renormalized Hamiltonian, 
describing the low energy quasiparticles, moving through a
self-consistently determined array of resonant scattering centers. 
Later,  we will see that 
the f-electron operators are composite objects, formed as 
bound-states between spins and conduction electrons. 
 
The mean-field Hamiltonian can be diagonalized in the form
\begin{equation}\label{}
H_{MFT}= \sum_{{\bk}\sigma }\left(a\dg _{{\bk}\sigma },b\dg
_{{\bk}\sigma } \right)\pmat{E_{{\bk}^{+}}&0\cr
0&E_{{\bk}^{-}}}
\pmat{a_{{\bk}\sigma }\cr b_{{\bk}\sigma }}
+N{\cal N}_{s}\left(\frac{\bar V V}{J}- \lambda q\right).
\end{equation}
Here
$a\dg _{{\bk}\sigma } = u_{\bk }c\dg _{\bk \sigma }+v_{\bk }f\dg_{\bk \sigma }
$ and $b\dg _{{\bk}\sigma } = - v_{\bk }c\dg_{\bk \sigma }+ u_{\bk }f\dg_{\bk \sigma }$
are linear combinations of $c\dg _{{\bk}\sigma }$ and $f\dg _{\bk\sigma} $, 
playing the role of 
``quasiparticle operators'' 
with corresponding energy eigenvalues
\begin{equation}\label{heavydeterminant}
{\rm  Det}
\left[E_{\bk}^{\pm}
\underline{1}-\pmat{\epsilon _{{\bk}}& V\cr
\bar V&\lambda}
 \right]= (E_{\bk \pm }-\epsilon_{\bk }) (E_{\bk \pm }- \lambda)- |V|^{2}= 0,
\end{equation}
or 
\begin{equation}\label{}
E_{{\bk}\pm } = \frac{\epsilon _{{\bk}}+\lambda}{2}\pm
\left[
\left(\frac{\epsilon _{{\bk}}-\lambda}{2}
\right)^{2}+ \vert V\vert ^{2} 
\right]^{\frac{1}{2}},
\end{equation}
and eigenvectors taking the 
BCS form 
\begin{eqnarray}\label{l}
\left\{ 
\begin{array}{c}
u_{\bk }\cr
v_{\bk }
\end{array}
\right\}= \left[\frac{1}{2}\pm \frac{(\epsilon_{\bk
}-\lambda)/2}{2\sqrt{
\left(\frac{\epsilon _{{\bk}}-\lambda}{2}
\right)^{2}+ \vert V\vert ^{2} 
}} \right]^{\frac{1}{2}}.
\end{eqnarray}
The hybridized dispersion
described by these energies is shown in Fig. \ref{fig20}.

Note that:

\begin{itemize}
\item 
The Kondo effect injects an f-band into the conduction sea, 
hybridizing with the conduction band to create two bands 
separated by a \underline{direct} ``hybridization gap'' of size $2V$
and a much smaller \underline{indirect} gap. If we put
$\epsilon_{\bk }=\pm D$, we see that the upper and lower edges of the
gap are given by 
\begin{equation}\label{}
E^{\pm } = \frac{\mp D +\lambda}{2}\pm  \sqrt{\left(\frac{\mp D-\lambda}{2}
\right)^{2}+V^{2}}\approx \lambda \pm  \frac{V^{2}}{D}, \qquad (D>>\lambda)
\end{equation}
so the indirect gap has a size 
$\Delta_{g}\sim 2V^{2}/D$, where $D$
is the half-bandwidth.  We will see shortly 
that 
$V^{2}/D \sim 
T_{K}$ is basically the single-ion Kondo temperature, so that 
that $V\sim \sqrt{T_{K }D}$ is the geometric mean of the band-width and
Kondo temperature. 

\item In the case when the chemical
potential lies in the gap, a \underline{Kondo} insulator is formed. 

\item A  conduction sea of electrons
has been transformed into a heavy Fermi sea of holes. 

\item The Fermi surface volume 
\underline{expands} in response to the formation of heavy electrons
(see Fig. \ref{chargefig}) to 
count the total
number of occupied quasiparticle states
\begin{equation}\label{}
N_{tot}= \langle \sum_{k\lambda \sigma }n_{k\lambda \sigma }
\rangle = \langle \hat n_{f}+\hat  n_{c}\rangle 
\end{equation}
where $n_{k\lambda \sigma }=a\dg _{k\lambda \sigma
}a_{k\lambda \sigma }
$ is the number operator for the quasiparticles and 
$n_{c}$ is the total number of conduction electrons. This means  
\begin{equation}\label{}
N_{tot}= N \frac{V_{FS}a^{3}}{(2\pi )^{3}}= Q + n_{c},
\end{equation}
where $a^{3}$ is the volume of the unit cell. This is rather
remarkable, for the expansion of the Fermi surface implies an
increased {\sl negative} charge density in the Fermi sea. Since charge is conserved,
we are forced to conclude there is a compensating $+Q|e|$ charge
density per unit cell provided by the Kondo singlets formed at each
site, as illustrated in Fig. \ref{chargefig}.

\end{itemize}

\fight= 4 truein
\fg{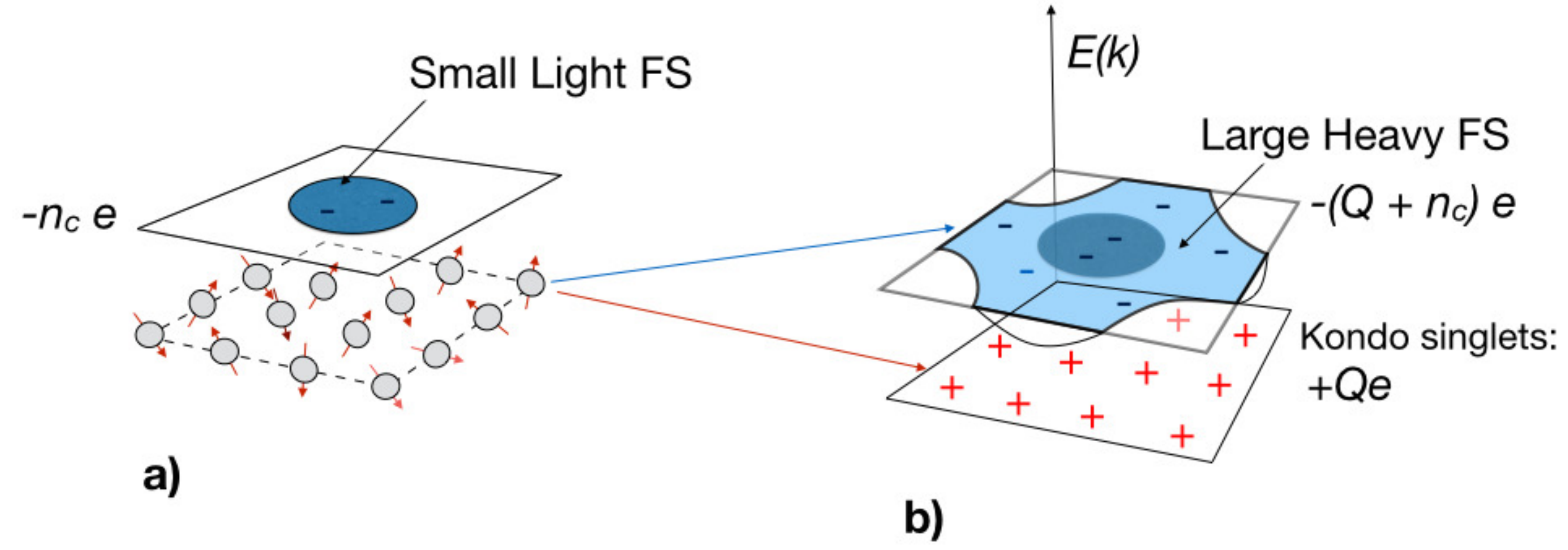}{
(a) High temperature state: small Fermi surface
with a background of spins; (b)Low temperature state where large Fermi
surface develops against a background of positive charge. Each spin
``ionizes'' into $Q$  heavy electrons, leaving behind 
a background of Kondo singlets, each  with charge $+Qe$.}{chargefig}
\fight=4 truein

\subsection{Free energy and Saddle Point}\label{}

Let us now use the results of the last section to 
calculate the mean-field free energy $F_{MFT}$ and determine,
self-consistently the parameters $\lambda$ and $V$ which set the
scales of the Kondo lattice.  
By diagonalizing the mean field Hamiltonian, we obtain 
\begin{equation}\label{}
\frac{F}{N} = -  T \sum_{\bk ,\pm}\ln \biggl[
1 + e^{-\beta E_{{\bk}\pm}} \biggr]
+ {\cal N}_{s}\left(\frac{V^{2}}{J}- \lambda
q\right).
\end{equation}

Let us discuss the ground-state, in which only the lower-band contributes to the
Free energy. As $T\rightarrow 0$, we can replace 
$-T\ln (1+e^{- \beta E_{\bk }})\rightarrow \theta (-E_{\bk })E_{\bk }$,
so  the ground-state energy $E_{0}= F (T=0)$ involves an integral over
the occupied states of the lower band:
\begin{equation}\label{}
\frac{E_{o}}{N{\cal N}_{s}}= \int_{-\infty }^{0} dE\rho ^{*} (E)
E + \left(\frac{ V^{2}}{J}- \lambda
q\right)
\end{equation}
where we have introduced the density of heavy electron 
states $\rho ^{*} (E)= \sum_{{\bk},\pm}\delta
(E-E^{(\pm)}_{{\bk}})$.
Now by (\ref{heavydeterminant}) the relationship between the energy
$E$ of the heavy electrons and the energy $\epsilon$ of the conduction electrons is 
\[
E= \epsilon + \frac{V^{2}}{E-\lambda }.
\]
As we sum over momenta $\bk $ within a given energy shell, 
there is a 
one-to-one correspondence between each conduction electron state and
each quasiparticle state, so we can write
$\rho^{*} (E) dE= \rho (\epsilon) d\epsilon $, where  the density 
of heavy
electron states 
\begin{equation}\label{dens2}
\rho^{*} (E)= \rho \frac{d\epsilon }{dE} = \rho \left(1 +
\frac{V^{2}}{(E-\lambda )^{2}} \right).
\end{equation}
Here we have approximated the underlying conduction electron density
of states by a constant $\rho =1/ (2D)$. 
The originally flat conduction electron density of states
is now replaced by a ``hybridization gap'', flanked by two sharp peaks
of width approximately $\pi \rho V^{2}\sim T_{K}$ (Fig. \ref{fig20}). Note that the lower
band-width is lowered by an amount $-V^{2}/D$.
With this information, we can carry out the integral over the energies,
to obtain
\begin{equation}\label{latticeenergy}
\frac{E_{o}}{N \Nsites}= \rho 
\int _{-D-V^{2}/D}^{0}dE  E
\left(1+ 
\frac{V^{2}}{(E-\lambda )^{2}}\right)+ \left(\frac{V^{2}}{J}- \lambda
q\right)
\end{equation}
where we have assumed that the upper band is empty, and the lower band
is partially filled. Carrying out the integral 
we obtain 
\begin{eqnarray}\label{l}
\frac{E_{o}}{N \Nsites}&=& 
-\frac{\rho }{2}\left(D+\frac{V^{2}}{D} \right)^{2}
+ 
\frac{\Delta }{\pi}
\int _{-D}^{0}dE \left(\frac{1}{E-\lambda} +
\frac{\lambda}{(E-\lambda)^{2}}\right)
+ \left(\frac{V^{2}}{J}- \lambda
q\right)
\cr
&=&
-\frac{D^{2}\rho }{2}
+ 
\frac{\Delta }{\pi}\ln \left(\frac{\lambda }{D} \right)
+
\left(\frac{V^{2}}{J}- \lambda
q\right)
\end{eqnarray}
where we have replaced $\Delta  =
\pi \rho V^{2}$, which is the width of an isolated f-resonance, and have 
dropped terms of order 
$O (\Delta^{2}/D)$. We can rearrange this expression, absorbing the
band-width $D$ and Kondo coupling constant into a single Kondo
temperature $T_{K}=De^{-\frac{1}{J\rho }}$ as follows
\begin{eqnarray}\label{18.130}
\frac{E_{0}}{N\Nsites} 
&=& -\frac{D^{2}\rho }{2}+
\frac{\Delta }{\pi}\ln \left(\frac{\lambda }{D} \right)
+
\left(\frac{\pi\rho V^{2}}{\pi \rho  J}- \lambda
q\right)
\cr
&=& -\frac{D^{2}\rho }{2}
+ 
\frac{\Delta }{\pi}\ln \left(\frac{\lambda }{D} \right)
+
\left(\frac{\Delta }{\pi \rho  J}- \lambda
q\right)
\cr
&=& -\frac{D^{2}\rho }{2}
+ 
\frac{\Delta }{\pi}\ln \left(\frac{\lambda }{De^{-\frac{1}{J\rho }}} \right)
- \lambda
q
\cr
&=& 
-\frac{D^{2}\rho }{2}
+ 
\frac{\Delta }{\pi}\ln \left(\frac{\lambda }{T_{K}} \right)
- \lambda
q.
\end{eqnarray}
This describes the energy of a family of Kondo
lattice models with different $J (D)$ and cutoff $D$, but fixed 
Kondo temperature. If we impose the constraint 
$\frac{\partial
E_{0}}{\partial \lambda }= \langle
n_{f}\rangle -Q=0$ we obtain $\frac{\Delta }{\pi \lambda} - q =0$,
so 
\begin{equation}\label{}
\frac{E_{o} (V)}{N \Nsites}= \frac{\Delta }{\pi }\ln \left(
\frac{\Delta }{\pi q e  T_{K}} 
\right)-\frac{D^{2}\rho }{2}, \qquad \qquad (\Delta = \pi\rho |V|^{2})
\end{equation}
\fight=2in
\fg{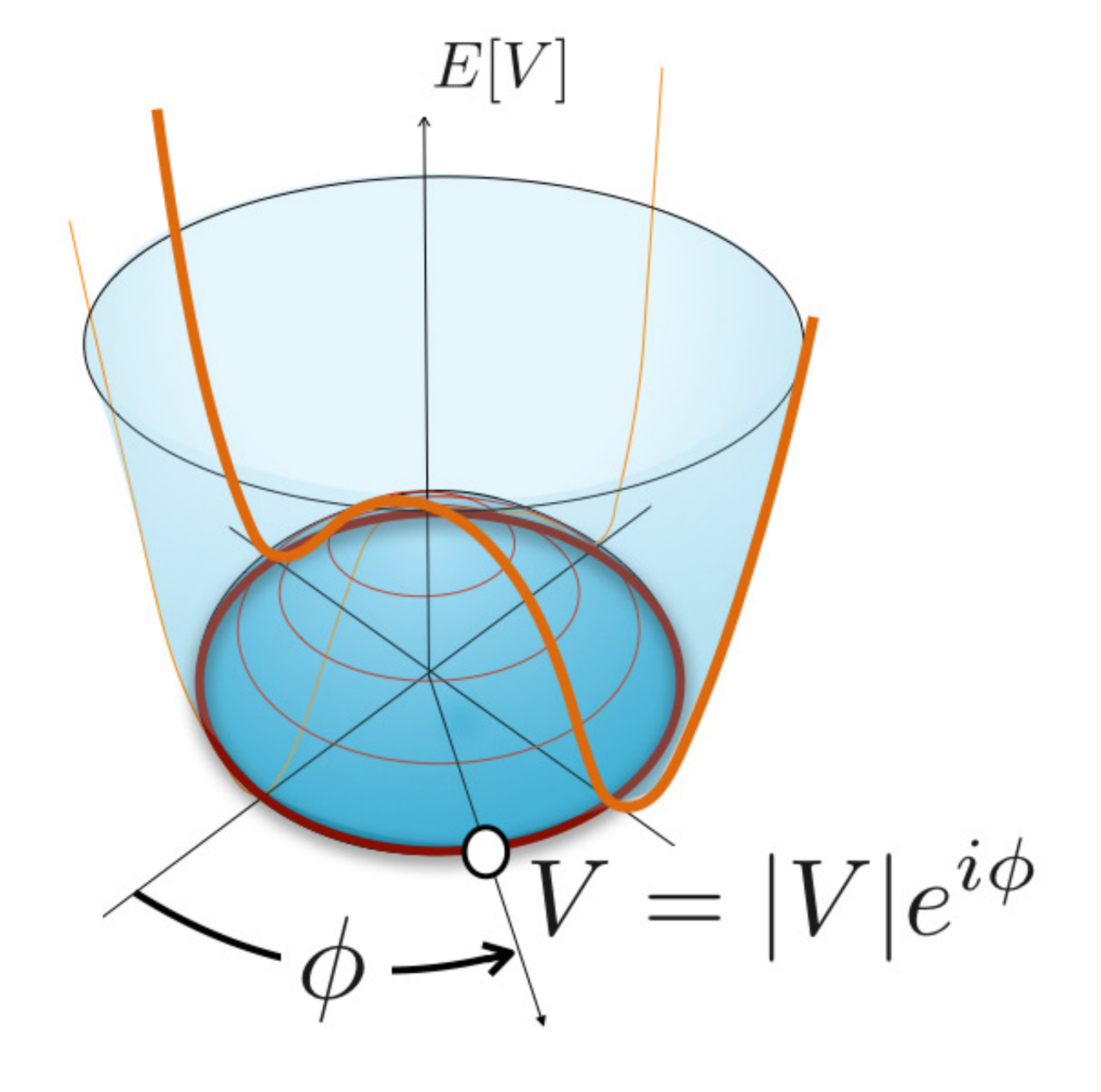}{Mexican hat potential for the
Kondo Lattice, evaluated at constant $\langle  n_{f}\rangle  =Q$ as a
function of a complex hybridization $V=|V|e^{i\phi }$}{figmex}
Let us pause for a moment to consider this energy functional
qualitatively. There are two points to be made

\begin{itemize}
\item 
The energy surface $E_{0} (V)$ is actually independent
of the phase of $V=|V|e^{i\phi }$(see Fig. \ref{figmex}), and has
the form of 
``Mexican Hat'' at low temperatures.   The minimum of this functional will
then determine a family of saddle point values $V= |V_{o}|e^{i\phi
}$, where $\phi $ can have any value. 
If we differentiate the ground-state energy with respect to $\Delta $,
we obtain
\[
0=\frac{1}{\pi }\ln \left(
\frac{\Delta  }{\pi q   T_{K}} 
 \right)
\]
or 
\[
\Delta =\pi  q  T_{K}
\]
confirming that $\Delta \sim T_{K}$.

\item The mean-field value of the constraint field $\lambda$ is
determined relative to the Fermi energy $\mu$. Were we to introduce a 
slowly varying 
external potential field to the conduction electron sea, then the chemical
potential becomes locally shifted so that $\mu \rightarrow \mu+e\phi (t) $,
So long as the field $\phi (t) $ is varied at a rate slowly compared with
the  Kondo temperature, the constraint field will always track with
the chemical potential, 
and since the constraint field is pinned to the chemical potential, 
$\lambda\rightarrow \lambda+ e\phi (t) $. In the process, the constraint
term will become 
\begin{equation}\label{}
\lambda (\hat n_{f} (j) -Q) \rightarrow \lambda (\hat n_{f} (j) -Q) + e\phi (t) (\hat n_{f} (j) -Q).
\end{equation}
Since the f-electrons now couple to the external potential $e\phi $
we have to ascribe a physical charge $e=-|e|$ to them. By contrast, the $-Q$ term
in the constraint must be interpreted as a ``background positive
charge'' $|e|Q\equiv |e|$ per site.  These lines of reasoning 
indicate that we should think of the Kondo effect as an {\bf
many-body ionization phenomenon} in which 
the neutral local moment splits up  into a negatively charged heavy
electron  and a stationary positive background
charge we can associate with the formation of a Kondo singlet.

\end{itemize}

\subsection{The Composite nature of the f-electron}\label{}

The matrix Green's function of the Kondo lattice reminds us of the
Nambu Green's function in superconductivity. It is given by 
\begin{equation}\label{}
{\cal G}_{\bk } (\tau ) = - \langle \psi_{\bk \sigma } (\tau )\psi\dg
_{\bk \sigma } (0)\rangle  \equiv \gmat{G_{c} (\bk ,\tau )
& G_{cf} (\bk,\tau )\cr
G_{fc} (\bk,\tau)& G_{f} (\bk,\tau )
}
\end{equation}
where $G_{c} (\bk ,\tau ) = - \langle c_{\bk } (\tau )c\dg _{\bk
\sigma } (0)\rangle $, $G_{cf } (\bk,\tau) = - \langle c_{\bk } (\tau
)f\dg_{\bk \sigma } (\tau )\rangle $ and so on. The anomalous off-diagonal
members of this Green's function remind us of the Gor'kov functions 
in BCS theory, and develop with the coherent hybridization. 
Using the two component notation (\ref{Ch16:twocomponent}), 
this Green's function can be written
\begin{equation}\label{}
{\cal G}_{\bk } (\tau )= - (\partial_{\tau }+ \underline{h}_{\bk
})^{-1}
\stackrel{\hbox{F.T.}}{
\xrightarrow{\hspace*{1.3cm}} }{\cal G}_{\bk } (i\omega_{n})=
(i\omega_{n}- \underline{h}_{\bk })^{-1},
\end{equation}
where F.T denotes a Fourier transform in imaginary time ($\partial_{\tau
}\rightarrow -i\omega_{n}$), 
or more explicitly,
\begin{eqnarray}\label{l}
{\cal G}_{\bk } (z) = (z- \underline{h}_{\bk })^{-1}
&=&\pmat{z- \epsilon_{\bk }& -V \cr - V &
z-\lambda}^{-1}
=
\pmat{G_{c} (\bk ,z) &G_{cf} (\bk ,z)
\cr
G_{fc} (\bk ,z)
&G_{f} (\bk ,z)
}
\cr
&=& \frac{1}{(z - \epsilon_{\bk }) (z-\lambda)- V^{2}}
\pmat{z- \lambda& V \cr
V & z-\epsilon_{\bk }}.
,\end{eqnarray}
where we have taken the liberty of analytically extending
$i\omega_{r}\rightarrow z$ into the complex plane. 
Now we can read off the Green's functions. In particular, the
``hybridized'' conduction electron Green's function is 
\begin{eqnarray}\label{Gccprop}
G_{c} (\bk ,z) &=& 
\parbox{15mm}{\bfyn(15,10)
\fmfleft{o}\fmfright{i}\fmf{heavy}{o,i}
\efyn}
= 
\frac{z-\lambda}{(z -
\epsilon_{\bk }) (z-\lambda)- V^{2}}\cr& =& \frac{1}{
z-\epsilon_{\bk }- \frac{V^{2}}{z-\lambda}
}\equiv \frac{1}{z-\epsilon_{\bk }- \Sigma_{c} (z)}
\end{eqnarray}
which we can interpret physically as 
conduction electrons scattering off resonant f-states at each site, 
giving rise to a momentum-conserving 
self energy 
\begin{equation}\label{}
\Sigma_{c} (z) = 
\parbox{40mm}{\bfyn(40,20)
\fmfset{arrow_len}{3mm}
\fmfleft{i}\fmfright{o}
\fmf{plain,tension=2}{i,a}
\fmf{dashes_arrow,tension=0.5,label=$O (1)$,label.side=right,label.dist=10}{a,b}
\fmfv{decor.shape=circle,decor.filled=full,fore=red,decor.size=3thick}{a}
\fmfv{label.angle=90,label=$V$}{a}
\fmfv{decor.shape=circle,decor.filled=full,fore=red,decor.size=3thick}{b}
\fmfv{label.angle=90,label=$V$}{b}
\fmf{plain,tension=2}{b,o}
\efyn} = \frac{V^{2}}{z-\lambda}.
\end{equation}
We see that the Kondo effect has injected a resonant scattering pole 
at energy $z=\lambda$ in the conduction electron self-energy.  This
resonant scattering lies at the heart of the Kondo effect.

\subsubsection{An absurd digression: the nuclear Kondo effect}\label{}

The appearance of this pole in the scattering raises a vexing question
in the Kondo effect: {\sl what is the meaning of the f-electron?}
This might seem like a dumb question, 
for in electronic materials the Kondo effect certainly involves 
localized f electrons, and surely, we can interpret this pole as
as the adiabatic renormalization of a hybridized band-structure.  This
is certainly true. 
Yet as purists, we do have to confess that our starting model, 
was a pure Kondo lattice model with only spin degrees of freedom:
they could even have been  \underline{nuclear} spins!  

This might seem absurd, yet nuclear spins
do couple antiferromagnetically with conduction electrons to produce
nuclear antiferromagnetism. 
Leaving aside practical issues of magnitude, 
we can learn something from the thought experiment in which the the nuclear spin
coupling to electrons {\sl is} strong enough to overcome the nuclear
magnetism. In this case, resonant bound-states would form with the
nuclear spin lattice giving rise to {\sl charged} heavy electrons,
presumably with an expanded Fermi surface.

From this line of argument we see that while it's tempting to
associate the heavy fermion with a physical f- or
d- electron localized inside the local moment, from a renormalization
group perspective, the heavy electron is an emergent excitation: a
fermionic bound-state formed between the conduction sea and the
neutral localized moments. This alternate point-of-view is useful, because
it allows us to contemplate the possibility of new kinds of Kondo
effect into states that are not adiabatically accessible from a band
insulator or metal.

\subsection{Cooper pair analogy}\label{}

There is a  nice analogy with superconductivity 
which helps to understand the composite nature of the heavy electron.
In a superconductor, electron pairs behave as loose composite
bosons described by the relation 
\begin{equation}\label{l}
\contraction{}{\psi}{_{\uparrow}(x)}{\psi}
\psi_{\uparrow}(x) \psi_{\downarrow} (x')  =  -F (x-x').
\end{equation}
Here $F (x-x')= -\langle T\psi _{\up} (1)\psi_{\dw} (2)\rangle $ is
the anomalous Gor'kov Greens function which determines the Cooper pair
wavefunction, extended over the 
coherence length $\xi\sim v_{F}/T_{c}$. 
A similar phenomenon takes place in the Kondo effect, but here the
bound-state develops between spins and electrons, forming a fermion,
rather than a boson. 
For the Kondo lattice, its perhaps more
useful to think in terms of a screening time
$\tau_{K} \sim \hbar / T_{K}$, rather than a length. 
Both the Cooper pair and heavy electron 
involve electrons that span decades of energy up to a cutoff, beit the 
Debye energy $\omega_{D}$ in superconductivity
or the (much larger) bandwidth $D$ in the Kondo effect
\cite{burdin,costimanini}.

To follow this analogy in greater depth,  recall 
that in  the path integral 
the Kondo interaction factorizes as
\begin{equation}\label{}
\frac{J}{N} 
c \dg _{\beta } S_{\alpha \beta }  c _{\alpha } 
\longrightarrow 
\bar V\left(c \dg _{\alpha  }f_{\alpha  } \right)
+
\left( f\dg _{\alpha }c_{\alpha } \right)V
 +N\frac{\bar VV}{J},
\end{equation}
so by comparing the right and left hand side, we see that the composite operators
$S_{ \beta \alpha } c _{\beta }$ and  $c\dg_{\beta }S_{\alpha \beta }$
behave as a single
fermion denoted by the contractions:\\
\sboxit{
\begin{equation}\label{thecomposite}
\frac{1}{N}\sum_{\beta }\contraction{}{S}{_{\beta\alpha  }  }{c}
S_{\beta\alpha  }  c _{\beta} 
= \left(\frac{\bar V }{J} \right)f_{\alpha },\qquad 
\frac{1}{N}\sum_{\beta }\contraction[2ex]{}{c\dg }{_{\beta}}{S}
c\dg_{\beta}S_{\alpha\beta }  
= \left(\frac{ V }{J} \right)f\dg _{\alpha },
\end{equation}
\rightline {\bf Composite Fermion}
}
\vskip 0.1truein
\noindent
Physically, this means that the spins bind high energy electrons, 
transforming themselves into composites which 
then hybridize with the conduction electrons. 
The resulting
``heavy fermions'' can be thought of as moments
ionized in the magnetically polar electron fluid to  form
mobile, negatively charged heavy electrons while  
leaving behind a positively charged ``Kondo singlet''.
Microscopically, the many body amplitude to scatter an electron off a
local moment develops a bound-state pole, which for large $N$
we can denote by the diagrams:
\begin{equation*}
\parbox{40mm}{\bfyn(40,18)
\fmfset{arrow_len}{3mm}
\fmfleft{i}\fmfright{o}
\fmftopn{t}{3}
\fmfbottomn{b}{3}
\fmf{fermion,tension=1.2}{i,a}
\fmf{dashes}{d,n}
\fmf{dashes,tension=3.0}{n,c,m}
\fmf{dashes}{m,a}
\fmf{fermion,tension=1.2}{d,o}
\fmf{phantom}{t2,h,c,l,b2}
\fmffreeze
\fmf{dashes_arrow}{d,n}
\fmf{dashes_arrow}{m,a}
\fmf{dashes_arrow,left=0.3}{a,h,d}
\fmf{fermion,right=0.3}{a,l,d}
\fmfv{decor.shape=circle,decor.filled=30,decor.size=16thick,label=$\Large \Gamma$,label.dist=0}{c}
\efyn}\equiv 
\parbox{40mm}{\bfyn(40,20)
\fmfset{arrow_len}{3mm}
\fmfleft{i}\fmfright{o}
\fmf{fermion}{i,a}
\fmf{dashes_arrow,tension=0.5,label=$O (1)$,label.side=right,label.dist=10}{a,b}
\fmfv{decor.shape=circle,decor.filled=full,fore=red,decor.size=3thick}{a}
\fmfv{label.angle=90,label=$V$}{a}
\fmfv{decor.shape=circle,decor.filled=full,fore=red,decor.size=3thick}{b}
\fmfv{label.angle=90,label=$\bar V$}{b}
\fmf{fermion}{b,o}
\efyn}
+
\parbox{40mm}{\bfyn(40,20)
\fmfset{arrow_len}{3mm}
\fmfleft{i}\fmfright{o}
\fmf{fermion}{i,a}
\fmf{dashes_arrow,tension=0.5}{a,b}
\fmf{fermion}{b,o}
\fmffreeze
\fmf{dashes_arrow,tension=0.5,label=$O (1/N)$,label.side=right,label.dist=10}{a,b}
\fmf{dbl_wiggly,left=0.5}{a,b}
\efyn}+\dots 
\end{equation*}
The leading diagram describes a kind of ``condensation'' of the
hybridization field; the second and higher terms describe the 
smaller $O (1/N)$ fluctuations around the mean-field theory. 

By analogy with superconductivity, 
we can associate a wavefunction associated with 
the temporal  correlations between spin-flips and conduction
electrons,  
as follows
\begin{equation}\label{thespinflip}
\frac{1}{N}\sum_{\beta }\contraction{}{c}{_{\beta } (\tau )}{S}
c _{\beta } (\tau ) S_{\beta \alpha } (\tau ')
= g (\tau -\tau ')\hat f_{\alpha } (\tau ').
\end{equation}
where the spin-flip correlation function 
$g(\tau -\tau ')$ is an analogue of the Gor'kov
function, extending over a coherence time $\tau_{K}\sim \hbar
/T_{K}$. Notice that 
in contrast to the Cooper pair,
this composite  object is a fermion and thus
requires a distinct operator $\hat f_{\alpha }$ for its expression.

\section{Heavy Fermion Superconductivity}\label{}
\index{heavy fermion superconductor}

We now take a brief look at heavy fermion superconductivity. 
There are a wide variety of heavy electron superconductors, almost all
of which are nodal superconductors, in which the pairing force derives
from the interplay of magnetism and electron motion. In the heavy
fermion compounds, as in many other strongly correlated electron
systems superconductivity
frequently develops at the border of magnetism, near the {\sl quantum
critical point} where the magnetic transition temperature has been
suppressed to zero. 
In some
of them, such as UPt$_{3}$ (T$_{c}$=0.5K)\cite{upt3} 
the superconductivity develops out of a well-developed heavy Fermi
liquid, and in these cases, we can consider the superconductor to be
paired by magnetic fluctuations within a well-formed heavy Fermi
liquid. 
However, in many other superconductors, such as
UBe$_{13}$(T$_{c}$=1K)\cite{ube13a,ube13b}, 
the 115 superconductors CeCoIn$_{5}$
(T$_{c}$=2.3K)\cite{sarrao}, CeRhIn$_{5}$ under
pressure (T$_{c}$=2K)\cite{tusonpark}, NpAl$_{2}$Pd$_{5}$(T$_{c}$=4.5K)\cite{Aoki:2007tn} and
PuCoGa$_{5}$ (T$_{c}$=18.5K)\cite{Sarrao:2002vh,curro}, the superconducting transition
temperature is comparable with the Kondo temperature. In many of these
materials, the entropy of condensation 
\begin{equation}\label{}
S_{c}= \int_{0}^{T_{c}}\frac{C_{V}}{T}dT
\end{equation}
can be as large as $( 1/3) R\ln 2$ per rare earth ion, indicating that
the spin is, in some-way entangling with the conduction electrons to
build the condensate. In this situation, we need to be able to
consider the Kondo effect and superconductivity on an equal footing.
\fight=6in
\fg{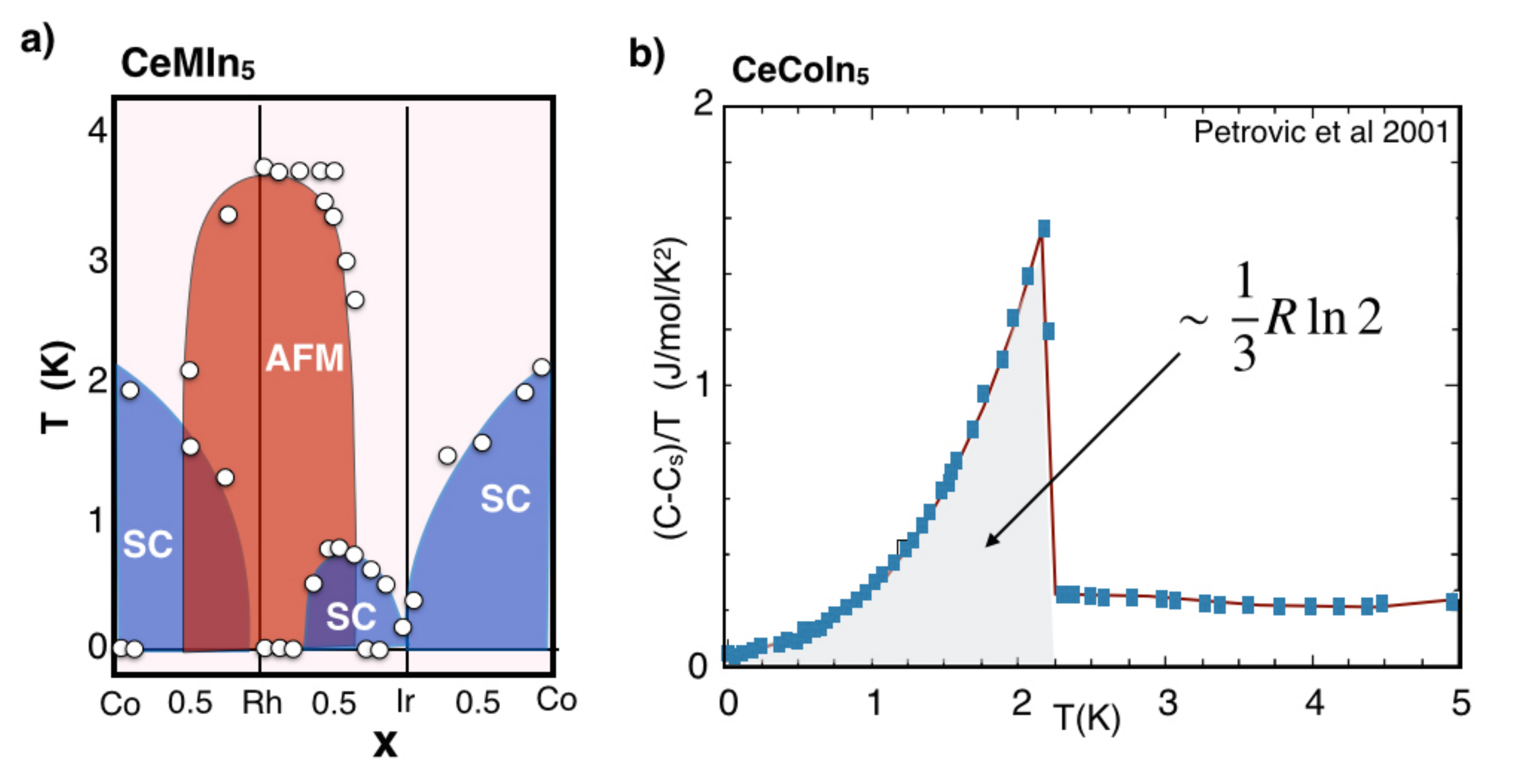}{(a) Phase diagram of 115 compounds
CeMIn$_{5}$, 
adapted
from \cite{sarraothompson07}, showing magnetic and superconducting
phases as a function of alloy concentration. 
(b) Sketch of specific heat coefficient of
CeCoIn$_{5}$, (with nuclear Schottky contribution subtracted), 
showing the large entropy of condensation associated with the
superconducting state. (After Petrovic et al 2001\cite{sarrao}). 
}{ceconin5}

\subsection{Symplectic spins and SP (N).}\label{}\index{SP(N)}
Although the SU(N) large $N$ expansion provides a very useful
description of the normal state of heavy fermion metals and Kondo
insulators, there is strangely, no superconducting solution. 
This short-coming lies in the very structure of the $SU (N)$
group. $SU (N)$  is perfectly tailored to particle physics, where the
physical excitations - the mesons and baryons appear as color
singlets, with the meson a 
a $q\bar q$ quark-antiquark singlet
while the baryon is an $N$-quark singlet $q_{1}q_{2}\dots q_{N}$,  
(where of course $N=3$ in reality). 
In electronic condensed matter, 
the meson becomes a particle-hole pair, but 
there are no two-particle singlets in $SU (N)$ beyond $N=2$. The
origin of this failure can be traced back to the absence of a
consistent definition of time-reversal symmetry in $SU (N)$ for
$N>2$. This means that 
singlet Cooper pairs and superconductivity can not develop at the large $N$ limit. 

A solution to this problem
which grew out an approach developed by Read and Sachdev\cite{sp2n}
for 
frustrated magnetism, is to use the symplectic 
group $SP (N)$\index{SP(N)}, where $N$ must be an even number\cite{Flint:2008fk,Flint:2010vz}. This little-known group
is a subgroup of $SU (N)$. 
In fact for $N=2$, $SU (2)=SP (2)$ are identical, but they diverge for
higher $N$. 
For example, 
$SU (4)$ has 15 generators, but its symplectic sub-group $SP(4)$ has only 10. 
At large $N$, $SP (N)$ has 
approximately half the number of generators of $SU (N)$.  
The symplectic property of the group allows it to consistently treat
time-reversal symmetry of spins and it also allows the formation of
two-particle singlets for any $N$.

One of the interesting aspects of $SP (N)$ spin operators, is their
relationship to pair operators.  Consider $SP (2)\equiv SU (2)$:
the pair operator is $\Psi \dg = f\dg_{\uparrow}f\dg_{\downarrow}$
and since this operator is a singlet, it commutes with the spin operators,
$[\Psi , \vec{S}]=[\Psi \dg ,\vec{S}]=0$
which, since $\Psi $ and $\Psi \dg $ are the generators of particle-hole
transformations, implies that the $SU(2)$ spin operator is
particle-hole symmetric.  It is this feature that is preserved by the
$SP (N)$ group, all the way out to $N\rightarrow \infty $. 
In fact, we can use
this fact to write down an $SP (N)$ spins as
follows: an $SU (N)$ spin is given by ${\cal S}^{SU (N)}_{\alpha \beta
}= f\dg_{\alpha }f_{\beta }$. Under a particle hole transformation
$f_{\alpha }\rightarrow {\rm Sgn} (\alpha )f_{-\alpha }$. 
If we take the particle-hold transform of the $SU (N)$
spin and add it to
itself we obtain  an $SP (N)$ spin, 
\boxit{\begin{equation}\label{}
S_{\alpha \beta  } = f\dg_{\alpha }f_{\beta } + {\rm Sgn} (\alpha
\beta )
f_{-\beta }f\dg _{-\alpha }, 
\end{equation}
\rightline{\bf Symplectic Spin operator}}
where the values of the spin indices are
$\alpha, \beta \in \{\pm 1/2, \dots,  \pm N/2 \}$.  
This spin operator commutes with the three isospin variables
\begin{equation}\label{}
\tau_{3}= n_{f}-N/2, \qquad \tau^{+ }= \sum_{\alpha >0}f\dg_{\alpha
}f\dg_{-\alpha }, \qquad \tau^{-}= \sum_{\alpha >0}f_{-\alpha }f_{\alpha }.
\end{equation}
With these local symmetries, the 
spin is continuous invariant under SU (2) particle-hole
rotations $f_{\alpha}\rightarrow
u f_{\alpha}+ v {\rm Sgn} \alpha f\dg _{-\alpha}
$, 
where $|u^{2}|+ |v^{2}|=1$, as you can verify.
To define an irreducible representation of the spin, we 
also have to impose a constraint on the Hilbert space, which in its
simplest form is $\tau_{3}=
\tau^{\pm }=0$, equivalent to $Q=N/2$ in the $SU (N)$ approach.  
In other-words, the s-wave part of the f-pairing must
vanish identically. 
\fight=4in
\fg{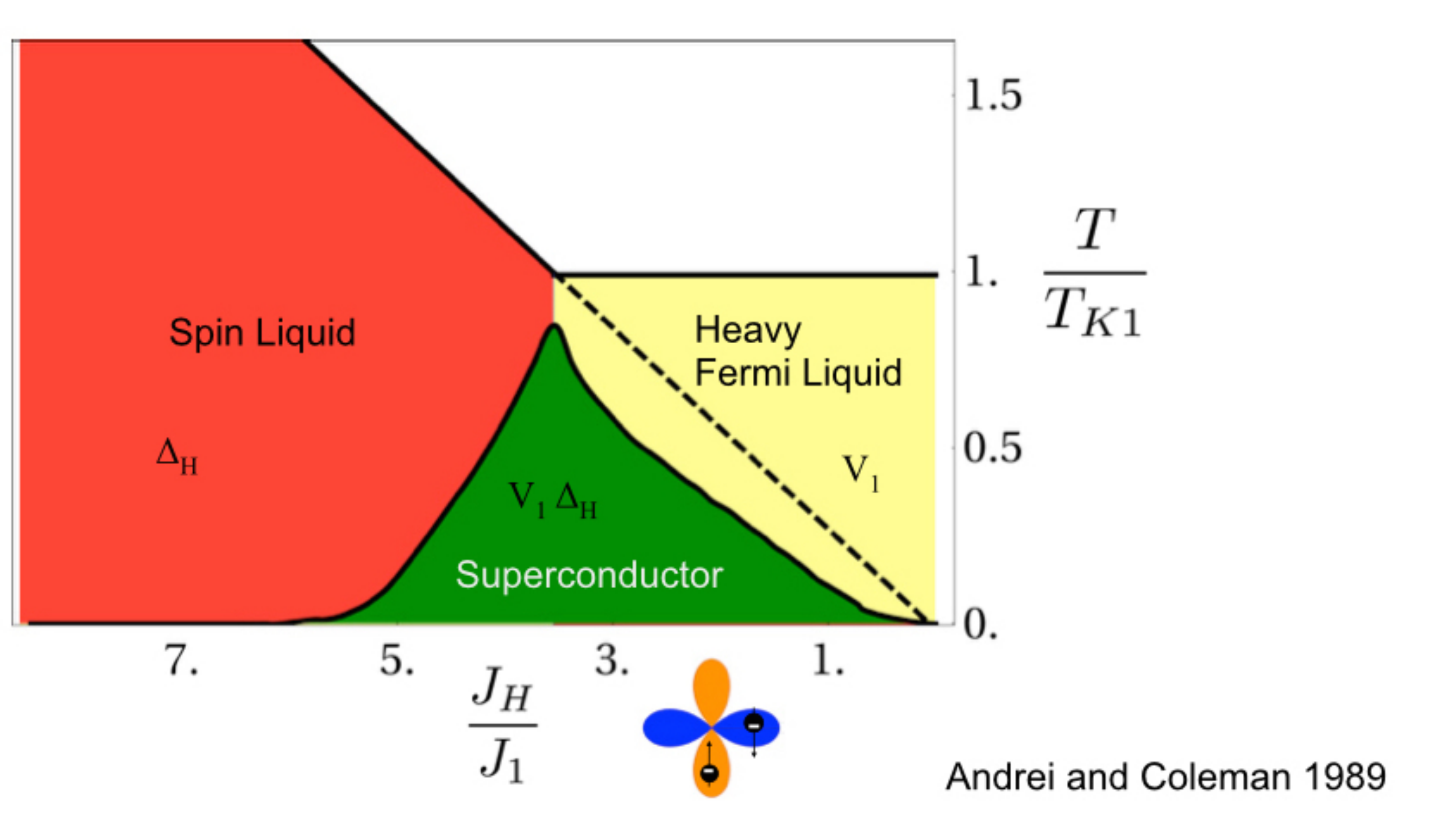}{Phase diagram for the two-dimensional
Kondo Heisenberg model, derived in the $SP (N)$ large $N$ approach,
adapted from \cite{colemanandrei}, courtesy Rebecca Flint.}{dwavehf}

\subsection{Superconductivity in the Kondo Heisenberg Model}\label{}

Let us take a look at the way this works in a nearest neighbor ``Kondo Heisenberg
model''\cite{colemanandrei},
\begin{equation}\label{kondoheisenberg}
H = H_{c}+H_{K}+ H_{M}.
\end{equation}
Here $H_{c}= \sum_{ \bk \sigma }\epsilon_{\bk  }c\dg_{\bk \sigma
}c_{\bk \sigma}$ describes the conduction sea, whereas $H_{K}$ and
$H_{M}$ are the Kondo and Heisenberg (RKKY)\index{RKKY interaction} interactions, respectively.
These take the form
\begin{eqnarray}\label{l}
H_{K} &=& \frac{J_{K}}{N}\sum_{j} c\dg_{j\alpha }c_{j\beta }S_{\beta
\alpha} (j)\rightarrow - \frac{J_{K}}{N}\sum_{i,j}\left(
(c\dg_{j\alpha}f_{j\alpha} )(f\dg _{j\beta}c_{j\beta})+ 
\tilde{\alpha }\tilde{\beta }
(c\dg_{j\alpha}f\dg _{j-\alpha} )(f _{j-\beta}c_{j\beta}) 
 \right)\cr
H_{M} &=& \frac{J_{H}}{2N} \sum_{(i,j)} S_{\alpha \beta } (j)S_{\beta
\alpha } (j)
\rightarrow - \frac{J_{H}}{N}\sum_{j}\left[
(f\dg_{i\alpha}f_{j\alpha} )(f\dg _{j\beta}f_{i\beta})+ 
\tilde{\alpha }\tilde{\beta}
(f\dg_{i\alpha}f\dg _{j-\alpha} )(f _{j-\beta}f_{i\beta}) 
 \right]
\end{eqnarray}
where we've introduced the notation $\tilde{\alpha }= {\rm Sgn}
(\alpha )$ and have shown how the interactions are expanded in 
into particle-hole and particle-particle channels. Notice how the
interactions are equally divided between particle-hole and
particle-particle channels. When we carry out the Hubbard Stratonovich
decoupling, in each of these terms, we obtain
\begin{eqnarray}\label{l}
H_{K}&\rightarrow &
\sum_j
\left[ c\dg_{j\alpha }\left(V_{j}f_{j\alpha}+
\tilde{\alpha }\Delta^{K}_{j}f\dg _{j-\alpha }\right) + 
{\rm H.c}\right] + 
N \left(
\frac{|V_{j}|^{2}+|\Delta^{K}_{j}|^{2} }{J_{K}} 
\right)
\cr
H_{H}&\rightarrow &\sum_{(i,j)}
\left[t_{ij}f\dg_{i\alpha }f_{j\alpha }+ \Delta_{ij}\tilde{\alpha
}f\dg_{i\alpha }f\dg_{j-\alpha }+{\rm H.c} \right] + N
\left[\frac{|t_{ij}|^{2}+ |\Delta_{ij}|^{2}}{J_{H}} \right]
\end{eqnarray}
At each site, we can always rotate the f-electrons in particle-hole
space to remove the ``Kondo pairing'' component and set $\Delta^{K}_{j}=0$, but the pairing terms
in the Heisenberg component can not be eliminated.  
This mean-field theory describes a kind of {\sl Kondo stabilized 
spin-liquid}\cite{colemanandrei}.  The physical picture is as follows: in practice, a
spin-liquid is unstable to magnetism, but 
its happy co-existence with the Kondo effect brings its energy below
that of the antiferromagnet. 
The hybridization of the $f$ with the conduction sea
converts the spinons of the spin-liquid into charged fermions. The
$t_{ij}$ terms describe various kind of exotic density waves. The
$\Delta_{ij}$ terms now describe pairing amongst the composite
fermions. 

To develop a simple theory of the superconducting state, 
we restrict our attention to uniform, static saddle points,
dropping the $t_{ij}$. Lets look at the resulting mean-field
theory. In two dimensions, this becomes 
\begin{equation}\label{}
H = \sum_{\bk , \alpha >0} 
(\tilde{c}\dg _{\bk  \alpha }, \tilde{f}\dg_{\bk  \alpha } )
\gmat{\epsilon_{\bk }\tau_{3}& V\tau_{1}\cr
V\tau_{1}& \vec{w}\cdot \vec{\tau }+\Delta_{H\bk }\tau_1
}\pmat{\tilde{c}_{\bk \alpha }\cr \tilde{f}_{\bk \alpha }} + {\cal
N}_{s}N \left(\frac{|V|^{2}}{J_{K}}+ 2\frac{|\Delta_{H}|^{2}}{J_{H}} \right)
\end{equation}
where 
\begin{equation}\label{}
\tilde{c}\dg _{\bk \alpha }= (c\dg_{\bk\alpha }, \tilde{\alpha }c_{-\bk ,
-\alpha }), \qquad 
\tilde{f}\dg _{\bk \alpha }= (f\dg_{\bk\alpha }, \tilde{\alpha }f_{-\bk ,
-\alpha })
\end{equation}
are Nambu spinors for the conduction and f-electrons.  The vector $\vec{W}$of
Lagrange multipliers couples to the isospin of the f-electrons:
stationarity of the Free energy with respect to this variable imposes
the mean-field constraint that $\langle \tilde{f}\dg  
\vec{\tau }f\rangle =0 $. The function
$\Delta_{H\bk } = \Delta_{\bk } (\cos k_{x}-\cos k_{y}) $ is the f-electron
pair wavefunction.  Here we've chosen a d-wave form-factor. For this
choice, the local f pair density automatically vanishes and so we need
only choose $\vec{w}= (0,0,\lambda)$, where $\lambda$ couples to
$\tau_3$ (imposing the constraint $n_{f}= N/2$). 
We could have also tried an extended 
extended s-wave pair wavefunction, but in this case, 
the induced s-wave pair density becomes finite, and the effect of the
$\vec{w}$
constraint is to suppress the transition temperature. 
By seeking stationary points in the free energy with respect to
variations in $\Delta_{H}$, $V$ and $\lambda$ one can derive the phase
diagram for d-wave pairing, 
shown in Fig. \ref{dwavehf}.  The mean-field theory shows that
superconductivity develops at the interface between the Fermi liquid
and the spin liquid.


\section{Topological Kondo Insulators}\label{}

One of the areas of fascinating development in the last few years, is the
discovery that Kondo insulators can develop {\sl topological order} to
form a {\sl Topological Kondo insulator}\index{topological Kondo insulator}
Topological order refers to the idea  that a quantum mechanical
ground-state can develop a non-trivial topology.  One of the defining
features of topological ground-states is the development of protected
surface states. The best known example of topological order 
the integer quantum Hall effect, where an integer filled Landau level
develops topological order that is
is responsible for the robust quantization of the Quantum Hall
effect\cite{Laughlin:1981jd,tknn,haldane88}.
In a remarkable series of discoveries in 2006, 
\cite{kanemele05,bernevig06,Moore:2006uk,rahulroy09,kane1,konig07,hsieh08,kane2}
it became clear that strong spin orbit coupling can play the role
of a synthetic magnetic field,  so that 
band insulators can also develop
a non-trivial topology while preserving time-reversal symmetry. Such
$Z_{2}$ topological band insulators\index{topological insulator} are 
defined by a single topological $Z_{2}= \pm 1$ index
that is positive in conventional insulators, but reverses in
topological $Z_{2}$ insulators.  This topological
feature manifests itself through the formation of robust conducting
surface states. 

In 2007,
Liang Fu and Charles Kane showed that if an insulator
has both time reversal and inversion symmetry 
\cite{kane2}, 
this $Z_{2}$ index is uniquely determined by the 
the parities $\delta_{in}$ of the Bloch states 
at the high symmetry points $\Gamma_{i}$ 
of the valence band which determine a ``$Z_{2}$ index''
\sboxit{
\begin{equation}\label{}
Z_{2}= \prod_{\Gamma_{i}}\delta (\Gamma_{i}) = \left\{
\begin{array}{cl}
+1&\hbox{conventional insulator}\cr
-1&\hbox{topological insulator}
\end{array} \right.
\end{equation}
\rightline{\bf Fu Kane formula for the $Z_{2}$ index of topological insulators}
}
where $\delta (\Gamma_{i})=\prod_{n} \delta_{in}$ 
is the  the product of the parities of the
of the occupied bands at the high-symmetry points
in the Brillouin zone. This formula allows one to determine 
whether an insulator  state is topological, merely by checking whether
the index $Z_{2}=-1$, without a
detailed knowledge of the ground-state wavefunction. 

It used to be thought that Kondo insulators could be regarded as
``renormalized silicon''. The discovery of topological
insulators\index{topological insulator} forced a re-evaluation of this
viewpoint. The large spin orbit coupling, and the odd-parity of the
f-states led to the proposal, by Dzero, Sun, Galitski and the author, 
\cite{Dzero2010} that Kondo insulators can become  topologically
ordered.
The Fu-Kane
formula has a special significance for Kondo insulators, which contain
{\sl odd parity} f-electrons hybridizing with {\sl even parity}
d-electrons. Each time an f-electron crosses through the band-gap,
exchanging with a conduction d-state, this changes the $Z_{2}$ index,
making it highly likely that certain Kondo insulators are
topological.  The oldest known Kondo insulator SmB$_{6}$, 
discovered almost 50 years ago was well known to possess a mysterious
low temperature conductivity plateau\cite{allen79,cooley}, and the
idea that this system might be a topological Kondo insulator provided
an exciting way of explaining this old mystery. 
The recent observation of robust~\cite{wolgasttki,kimtki} conducting
surface states in the oldest Kondo insulator \SMB  supports
one of the key elements of 
this  prediction, prompting a revival of interest 
in Kondo insulators as a new route for studying 
the interplay of strong interactions and topological order. 

SmB$_{6}$ is really a mixed valent system, which takes us a little
beyond the scope of this lecture. One of the other issues with
SmB$_{6}$, is that its local crystal field configuration is likely to
be a $\Gamma_{8}$ quartet state\cite{Alexandrov13}, rather than a Kramers
doublet. Nevertheless, key elements of its 
putative topological Kondo insulating state
are nicely illustrated by a {\sl spin-orbit coupled} Kondo-Heisenberg
model, describing the interaction of Kramer's doublet f-states with a
d-band. The model is essentially 
identical with  (Eq. \ref{kondoheisenberg})
\begin{equation}\label{}
H= \sum_{ \bk \sigma }\epsilon_{\bk  }\psi\dg_{\bk \sigma
}c_{\bk \sigma} +  {J_{K}}\sum_{j} \psi\dg_{j\alpha }\psi_{j\beta }S_{\beta
\alpha} (j)+{J_{H}} \sum_{i,j} S_{\alpha \beta } (i)S_{\beta
\alpha } (j)
\end{equation}
with an important modification that takes into account the large
spin-orbit coupling and the odd-parity of the f-states. This forces
the local Wannier states $\Psi_{j\alpha }$ that exchange spin with the
local moment to be odd parity combinations of nearest neigbour
conduction electrons, given by
\begin{equation}\label{}
\psi\dg_{j\alpha} = \sum_{i, \sigma } c\dg_{i\sigma}\Phi_{\sigma \alpha } (\bR_{i}-\bR_{j})
\end{equation}
We'll consider a simplified model with the form factor 
\newcommand{\bs}{{\bf s}}
\begin{equation}\label{}
\Phi (\bR )=  \left\{\begin{array}{cc}
-i \hat  R\cdot \frac{\vec{\sigma }}{2}
, & \bR \in \hbox{n.n}\cr
0 & \hbox{otherwise}
\end{array}
\right.
\end{equation}
This form factor describes the spin-orbit 
mixing between states with orbital angular
momentum $l$ differing by one, such as $f$ and $d$ or
$p$ and $s$ orbitals. 
The odd-parity of the form-factor 
$ \Phi (\bR) = - \Phi (-\bR)$ 
derives from the odd-parity $f$- orbitals, while
the prefactor $-i$ ensures that the hybridization
is invariant under time-reversal. 
The Fourier transform of
this Form factor, $\Phi (\bk ) = \sum_{\bR}\Phi (\bR)e^{i \bk\cdot
\bR}$ 
 is then 
\begin{equation}\label{}
\Phi (\bk ) = \vec{s}_{\bk }\cdot \vec{\sigma }
\end{equation}
where  the s-vector $ \vec{s}_{\bk } = (\sin k_{1},\sin k_{2},\sin k_{3})$ is the 
periodic equivalent of  the unit momentum vector
$\hat \bk $. Notice how $\vec s ({\Gamma_{i}})=0$  vanishes at the high
symmetry points.  

The resulting mean-field Hamiltonian takes the form 
\begin{equation}\label{}
H_{TKI} = \sum_{\bk} \psi\dg_{\bk}h (\bk )\psi_{\bk } + {\cal
N}_{s}\left[
\left(\frac{V^{2}}{J_{K}} + \frac{3t^{2}}{J_{H}}- \lambda Q\right)
 \right]
\end{equation}
where $\psi\dg_{\bk} = (c\dg _{\bk \sigma }, f\dg_{\bk \sigma})$ and 
\begin{equation}\label{}
h (\bk )= \pmat{\epsilon_{\bk }& V \vec{\sigma }\cdot
\vec{s}_{\bk}\cr
V\vec{\sigma }\cdot\vec{s}_{\bk}& \epsilon_{f\bk }
}
\end{equation}
while $\epsilon_{f\bk }= 2 t_{f} (c_{x}+c_{y}+c_{z})+ \lambda $ ($c_{l}\equiv \cos k_{l}$) is the
dispersion of the f-state resulting from a mean-field decoupling of
the intersite Heisenberg coupling in the particle-hole channel. 
For small $\bk $, the hybridization in  Hamiltonian $h (\bk )$ takes
the form $V \vec{\sigma}\cdot \bk $, a form which 
closely resembles the topologically non-trivial 
triplet p-wave gap structure of 
superfluid He-3B. 
Like He-3B, the
hybridization only develops at low temperatures, making SmB$_{6}$ an
adaptive insulator.
\fg{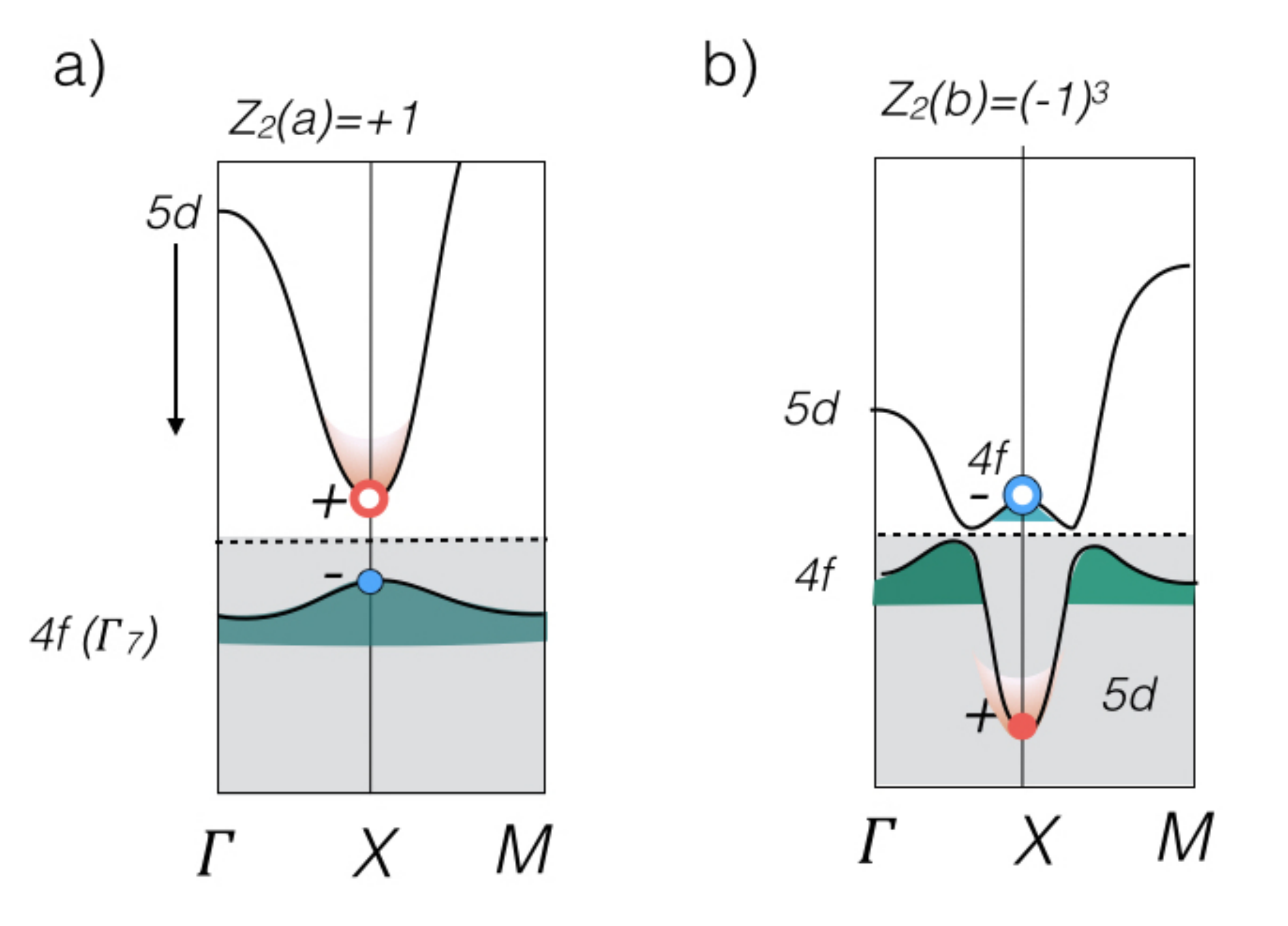}{(a) When the d-band is above the filled
f-band, a trivial insulator is formed. (b) When the d-band crosses the
f-band at the three X-points, the $Z_{2}$ parity changes sign, giving
rise to a topological insulator.}{bandcrossing}

Let us for the moment treat $h (\bk )$ as a rigid band structure. 
Suppose the f-band were initially
completely filled, with a completely empty d-band above it. (See
Fig. \ref{bandcrossing} a).  This situation corresponds to a
conventional band insulator with $Z_{2}=+1$ Next, let us lower the d-conduction
band until the two bands cross at a high symmetry point, causing the
gap to close, and then to re-open. 
We know, from dHvA studies of the iso-electronic
material LaB$_{6}$\cite{Fisk:1976tg} (whose band-structure is identical to SmB$_{6}$ but lacks the magnetic
f-electrons), from ARPES 
studies\cite{Neupane:2013hz,,PhysRevB.88.121102,Denlinger:2013bj}, that 
in 
SmB$_{6}$, the d-band crosses through the Fermi surface at 
at the three $X$ points. 
Once the d-band is lowered through
the f-band around the three $X$ points, 
the odd-parity f-states at the $X$ point move up into the conduction
band, to be replaced by even-parity d-states.  This changes the sign
of $Z_{2}\rightarrow (-1)^{3}=-1$, producing a topological
ground-state.  Moreover, since there are three crossing, we expect
there to be three spin-polarized surface Dirac cones. 

We end by noting that at the time of writing, our understanding of the
physics SmB$_{6}$ is in  rapid flux on both the experimental and
theoretical front.
Spin resolved ARPES\cite{Xu:1jv}
measurements have detected the presence of spin-textures in the
surface Fermi surfaces around the surface $\bar X$ point, a strong
sign of topologically protected surface states. Two recent theoretical
works\cite{Legner:2015tt,Baruselli:2015wq} have shown that the spin textures seen in these experiments are
consistent with a spin-quartet ground-state in SmB$_6$.
Despite this progress, consensus on the topological nature of
SmB$_{6}$
has not yet been achieved, and 
competing groups have offered alternate interpretations
of the data, including the possibility of polarity-driven surface
metallicity\cite{Zhu13} and 
and Rashba-split
surface states, both,  of a non-topological origin. 
\cite{Hlawenka:2015wj}.  Another area of experimental controversy concerns the
possible de-Haas van Alphen oscillations created by surface
topological excitations, with one report of the detection of surface
de Haas van Alphen signals\cite{Li:2014hk} and a recent, very
remarkable report of {\sl bulk} de Haas van Alphen signals associated with
unhybridized, quantum critical d-electrons\cite{suchitradhva}.

\section{Co-existing magnetism and Kondo effect}\label{}

In this short lecture, I've given a quick introduction to the 
paramagnetic phases of heavy fermion systems. One of the 
of major open questions in heavy fermion and Kondo lattice
physics concerns the physics of magnetism, and the right way to
describe the development of magnetism within these materials.
There is growing evidence that magnetism and the Kondo effect can
co-exist, sometimes homogeneously, and sometimes inhomogeneously. 
For example, 
 In the 115 superconductor CeRhIn$_{5}$ there is evidence for a 
microscopic and homogeneous coexistence of local moment 
magnetism and heavy fermion superconductivity under pressure \cite{Kne};
By contrast, 
in the geometrically frustrated CePdAl \cite{CePdAl1,CePdAl2}, two
thirds of the Cerium sites spontaneously develop magnetism,
leaving the other third to undergo a Kondo effect \cite{Fri2}. What is
the right way to describe these co-existent states?

One possibility that I have worked on with Aline Ramires\cite{susy1,alinenew}
is the use of a {\sl supersymmetric}
is the use of a ``supersymmetric'' spin representation of the spin
\begin{equation}\label{}
S_{\alpha \beta } = f\dg_{\alpha }f_{\beta }+ b\dg_{\alpha }b_{\beta }
\end{equation}
where the $f\dg_{\alpha }$ and $b\dg_{\alpha }$ are fermionic and
bosonic creation operators. Such a representation permits in
principle, the existence of ``two fluid'' ground-states, involving a
Gutzwiller projection of bosonic and fermionic wavefunctions
\begin{equation}\label{}
\vert \Psi \rangle  = P_{G} 
\vert \Psi_{F}\rangle \vert
\Psi_{B}\rangle, 
\end{equation}
where $\vert \Psi_{F}\rangle $ is the fermionic component of the wavefunction
describing the Kondo quenched local moments while $\vert
\Psi_{B}\rangle $ 
describes the formation of long-range magnetic correlations within a
bosonic RVB wavefunction, while 
\begin{equation}\label{}
P_{G }= \int \prod_{j} \frac{d\theta_{j}}{2\pi} e^{i \theta_{j} (n_{B}+n_{F}-1)}
\end{equation}
is a Gutzwiller projection operator onto the state with one spin per
site.  We have been trying to describe such mixed state wavefunctions
in the large-$N$ limit, seeking saddle point solutions where a bosonic
and fermionic fluid co-exist\cite{alinenew}. 
One of the ideas that emerges from this kind of approach, is the
possibility that the soft modes at a Quantum Critical point might develop
fermionic character, a kind of emergent supersymmetry\cite{susy2}.

\noindent {\bf Acknowledgments}

This article was written with the support of the National Science
Foundation under Grant No. NSF DMR-1309929.

\end{fmffile}
\clearchapter

\bibliographystyle{correl}
\bibliography{kondo21v2}

\end{document}